\documentclass[twocolumn]{aastex62}

\received{2018 August 6}
\revised{2018 August 17}
\accepted{2018 August 27}
\submitjournal{ApJ}

\shorttitle{An Ultra Metal-poor Star Near the Hydrogen-burning Limit}
\shortauthors{Schlaufman et al.}

\begin{document}

\title{An Ultra Metal-poor Star Near the Hydrogen-burning
Limit\footnote{This paper includes data gathered with the 6.5 meter
Magellan Telescopes located at Las Campanas Observatory, Chile.}}

\correspondingauthor{Kevin C. Schlaufman}
\email{kschlaufman@jhu.edu}

\author[0000-0001-5761-6779]{Kevin C. Schlaufman}
\affiliation{Department of Physics and Astronomy \\
Johns Hopkins University \\
3400 North Charles Street \\
Baltimore, MD 21218, USA}

\author{Ian B. Thompson}
\affiliation{Carnegie Observatories \\
813 Santa Barbara Street \\
Pasadena, CA 91101, USA}

\author[0000-0003-0174-0564]{Andrew R. Casey}
\affiliation{School of Physics \& Astronomy \\
Monash University \\
Clayton 3800, Victoria, Australia}
\affiliation{Faculty of Information Technology \\
Monash University \\
Clayton 3800, Victoria, Australia}

\begin{abstract}

\noindent
It is unknown whether or not low-mass stars can form at low metallicity.
While theoretical simulations of Population III (Pop III) star formation
show that protostellar disks can fragment, it is impossible for those
simulations to discern if those fragments survive as low-mass stars.
We report the discovery of a low-mass star on a circular orbit with
orbital period $P = 34.757 \pm 0.010$ days in the ultra metal-poor (UMP)
single-lined spectroscopic binary system 2MASS J18082002--5104378.
The secondary star 2MASS J18082002--5104378 B has a mass $M_{2} =
0.14_{-0.01}^{+0.06}~M_{\odot}$, placing it near the hydrogen-burning
limit for its composition.  The 2MASS J18082002--5104378 system is
on a thin disk orbit as well, making it the most metal-poor thin
disk star system by a considerable margin.  The discovery of 2MASS
J18082002--5104378 B confirms the existence of low-mass UMP stars and its
short orbital period shows that fragmentation in metal-poor protostellar
disks can lead to the formation and survival of low-mass stars.  We use
scaling relations for the typical fragment mass and migration time
along with published models of protostellar disks around both UMP and
primordial composition stars to explore the formation of low-mass Pop
III stars via disk fragmentation.  We find evidence that the survival
of low-mass secondaries around solar-mass UMP primaries implies the
survival of solar-mass secondaries around Pop III primaries with masses
$10~M_{\odot} \lesssim M_{\ast} \lesssim 100~M_{\odot}$.  If true,
this inference suggests that solar-mass Pop III stars formed via disk
fragmentation could survive to the present day.

\end{abstract}

\keywords{binaries: spectroscopic --- Galaxy: disk --- stars: formation ---
stars: low-mass --- stars: Population II --- stars: Population III}

\section{Introduction}

The first stars in the universe---the so called Population III (Pop III)
stars\footnote{The current state of Pop III star formation research
is reviewed in \citet{bro13}, \citet{glo13}, and \citet{gre15}.}---are
unique in that they are composed exclusively of the stable products of
big bang nucleosynthesis: hydrogen, helium, and a dusting of lithium
\citep[e.g.,][]{alp48,cyb16}.  Unlike metal-enriched gas that efficiently
cools via dust and metal-line emission, metal-free gas can only cool
significantly via atomic (H), molecular (H$_{2}$), and deuterated (HD)
hydrogen emission.  Hydrogen can only cool gas down to temperatures $T
\lesssim 10^{4}$ K, and H$_{2}$ is a poor coolant at $T \lesssim 200$ K.
While HD can cool gas below $T \approx 200$ K, the small cosmological
ratio of deuterium to hydrogen limits its contribution to cooling.
The net result is that primordial composition gas cannot efficiently
cool, and this inefficient cooling implies large Jeans masses and
therefore only massive star formation.  Both simple models and the
earliest cosmologically self-consistent three-dimensional hydrodynamic
simulations of Pop III star formation supported this picture and suggested
a characteristic Pop III star mass $M_{\ast} \gtrsim 100~M_{\odot}$
\citep[e.g.,][]{sil83,teg97,bro99,bro02,abe00,abe02,yos06,osh07}.
These first-generation hydrodynamic simulations also indicated that Pop
III stars should form as single stars in isolation.

These first-generation simulations did not include the effect of
radiative feedback from the forming star.  It has subsequently
been shown that radiative feedback plays a critical role in halting
accretion onto a Pop III star and consequently in setting its final
mass \citep[e.g.,][]{mck08,hos11,sta12,sus13,hos16}.  The most recent
simulations predict that Pop III stars should form over a range in mass
$10~M_{\odot} \lesssim M_{\ast} \lesssim 1000~M_{\odot}$.  This wide range
in mass is a result of the varying far-ultraviolet background due to the
expansion of the universe and the change in the characteristic mass of Pop
III stars with time \citep[e.g.,][]{hir14,hir15,sus14}.  At the low-mass
end, these theoretical predictions are supported by nucleosynthesis
calculations that suggest that the relative abundances of metals observed
in the most metal-poor stars are best explained by the supernovae of stars
with initial masses in the range $10~M_{\odot} \lesssim M_{\ast} \lesssim
100~M_{\odot}$ \citep[e.g.,][]{tum06,tak14,tom14,pla15,deb17,fra17,ish18}.
The abundances expected from the pair-instability supernovae of Pop III
stars with $M_{\ast} \gtrsim 100~M_{\odot}$ are rarely seen, and this may
suggest that the characteristic mass scale of Pop III stars is $M_{\ast}
\lesssim 100~M_{\odot}$ \citep[e.g.,][]{aok14,tak18}.

While the observed chemical abundances of metal-poor stars can
be used to constrain the properties of massive Pop III stars
that explode as supernovae, they have nothing to say about the
existence of low-mass Pop III stars.  This is a major problem, as
fragmentation at the molecular core and protostellar disk scales
has now been seen in numerical simulations of Pop III star formation
\citep[e.g.,][]{cla08,cla11a,cla11b,tur09,sta10,sta12,sta16,gre11,gre12,dop13,sta13,sta14,hir17,ria18}.
Whereas fragmentation at the molecular core scale will likely lead to
massive binary stars, the emergence of gravitationally bound solar-mass
clumps in protostellar disks via gravitational instability has the
potential to produce low-mass Pop III stars that may be observable in
the Milky Way.

Even though the reality of fragmentation is now widely accepted, it
is impossible to run these simulations forward in time long enough
to evaluate whether or not these fragments will survive as low-mass
stars or migrate inward and merge with the primary forming at the
center of the system \citep[e.g.,][]{vor13}.  Consequently, it is
still unknown whether or not low-mass Pop III stars ever existed.
Despite decades of searching, no low-mass primordial composition
star has ever been found.  While it is possible to set upper limits
on the occurrence of low-mass Pop III stars in the Milky Way, the
heterogeneous nature and singular focus on the most metal-poor objects
in the course of most metal-poor star surveys render them very uncertain
\citep[e.g.,][]{sal07,har15}.  It is also possible that the accretion of
gas, dust, or even asteroids from the interstellar medium may contaminate
the atmosphere of a Pop III star such that it appears to be an extreme
Pop II star \citep[e.g.,][]{fre09,joh15,kom15,kom16,she17,tan17,tan18}.
This ambiguity illustrates the importance of methods that do not rely on
the abundances of metal-poor stars to evaluate whether or not low-mass
Pop III stars exist.

As we will show, the occurrence and properties of low-mass objects around
the most metal-poor stars provide a chemical-abundance-independent way
to investigate low-mass Pop III star formation.  We report the discovery
of a low-mass ultra metal-poor (UMP)\footnote{Following the definition
of \citet{bee05}, UMP objects have $-5.0 \lesssim \mathrm{[Fe/H]}
\lesssim -4.0$.} star in the single-lined spectroscopic binary
system 2MASS J18082002--5104378.  The secondary in the system, 2MASS
J18082002--5104378 B, has a mass $M_{2} = 0.14_{-0.01}^{+0.06}~M_{\odot}$,
near the hydrogen-burning limit for its metallicity.  Because of its
low mass and metallicity, it has the fewest grams of heavy elements
of any known star.  We use models of protostellar disks around both
UMP and Pop III protostars plus scaling relations for the fragment
mass and migration time to argue that the existence of the low-mass
UMP star 2MASS J18082002--5104378 B and the extremely metal-poor
(EMP)\footnote{Following the definition of \citet{bee05}, EMP objects have
$-4.0 \lesssim \mathrm{[Fe/H]} \lesssim -3.0$.} brown dwarf HE 1523--0901
B discovered by \citet{han15} implies the survival of solar-mass
fragments around Pop III stars in the mass range $10~M_{\odot} \lesssim
M_{\ast} \lesssim 100~M_{\odot}$.  We describe our observations of the
2MASS J18082002--5104378 system and our data reduction in Section 2.
We detail our radial velocity measurement, Keplerian orbital analysis,
and Galactic orbit determination of 2MASS J18082002--5104378 and then
derive the significance of its existence for Pop III star formation in
Section 3.  We discuss the overall results and implications in Section
4 and conclude by summarizing our findings in Section 5.

\section{Observations}

The UMP nature of the primary in the 2MASS J18082002--5104378 system was
discovered by \citet{mel16}.  Those authors measured the spectroscopic
stellar parameters we reproduce in Table~\ref{tbl-1}.  \citet{mel16}
also noted that 2MASS J18082002--5104378 was a single-lined spectroscopic
binary with radial velocity variations in the most luminous component on
the order of 100 km s$^{-1}$ over a five month period from 2014 October
to 2015 March.  In addition, \citet{mel16} found no evidence for a second
set of absorption lines in their high-resolution, high signal-to-noise
ratio (S/N) spectra.  These latter two facts attracted our attention
to the 2MASS J18082002--5104378 system, as radial velocity variations
of 100 km s$^{-1}$ over a five month interval with no evidence for a
second set of lines in high-quality high-resolution spectra could only
be explained by a neutron star or stellar mass black hole.

\begin{deluxetable}{lcl}
\tablecaption{Spectroscopic Stellar Parameters for 2MASS
J18082002--5104378 A from \citet{mel16}\label{tbl-1}}
\tablewidth{0pt}
\tablehead{
\colhead{Property} & \colhead{Value} & \colhead{Units}}
\startdata
Effective temperature $T_{\mathrm{eff}}$ & $5440 \pm 100$ & K \\
Surface gravity $\log{g}$ & $3.0 \pm 0.2$ & \\
Microturbulence $v_{t}$ & $1.5 \pm 0.2$ & km s$^{-1}$ \\
Metallicity [Fe/H] & $-4.07 \pm 0.07$ & \\
\enddata
\end{deluxetable}

We subsequently observed the 2MASS J18082002--5104378 system 14 times
between 2016 April and 2017 July with the Magellan Inamori Kyocera Echelle
(MIKE) spectrograph on the Magellan Clay Telescope at Las Campanas
Observatory \citep{ber03,she03}.   We used the 0\farcs7~slit and the
standard blue and red grating azimuths, yielding spectra between 332
nm and 915 nm with resolution $R \approx 41,\!000$ in the blue and $
R \approx 35,\!000$ in the red.  Exposure times between two and five
minutes depending on conditions produced spectra that have $\mathrm{S/N}
\approx 25$ pixel$^{-1}$ at 400 nm and $\mathrm{S/N} \approx 40$
pixel$^{-1}$ at 460 nm.  On each night, we also observed the radial
velocity standards HIP 81294 or HIP 90522 from \citet{sou13}.  Exposures
on both the 2MASS J18082002--5104378 system and the radial velocity
standards were followed immediately by ThAr lamp spectra.  We collected
all other calibration data (e.g., bias, quartz \& ``milky" flat field,
and additional ThAr lamp frames) in the afternoon before each night of
observations.  We reduced the raw spectra and calibration frames using the
\texttt{CarPy}\footnote{\url{http://code.obs.carnegiescience.edu/mike}}
software package \citep{kel00,kel03}.

In parallel, between 2016 June and 2017 July we collected 31 epochs of
low-resolution spectroscopy with the Gemini Multi-Object Spectrograph
(GMOS) on the Gemini South telescope \citep{hoo04,gim16}.  We used
the 0\farcs5~slit and the B600 grating with a central wavelength of
500 nm, producing spectra between 350 and 650 nm with resolution $R
\approx 1,\!700$.  For each epoch, we collected three individual spectra
each with an exposure time 180 s yielding $\mathrm{S/N} \approx 100$
pixel$^{-1}$ at 430 nm.  After each science exposure, we obtained CuAr
lamp spectra and quartz--tungsten--halogen flat fields.  We also observed
the radial velocity standard HIP 81294 with the same set-up.  We debiased
each exposure using its overscan region.  We flat fielded and wavelength
calibrated each science frame using the quartz--tungsten--halogen and CuAr
frames taken immediately following the science exposure.  We propagated
data quality and inverse variance arrays at every step to obtain at the
end a Gaussian extraction of the wavelength-calibrated and transformed
image.  We continuum normalized each spectrum using a spline function
with a mask that excluded strong absorption lines and the gaps between
detector chips.

\section{Analysis}

\subsection{Radial Velocity Measurement}

We measured radial velocities from our high-resolution spectra using
\citet{ton79} one-dimensional Fourier cross-correlation as implemented
in the \texttt{FXCOR} task available in the Image Reduction and Analysis
Facility \cite[IRAF;][]{tod86,tod93}.\footnote{IRAF is distributed by
the National Optical Astronomy Observatories, which are operated by
the Association of Universities for Research in Astronomy, Inc., under
cooperative agreement with the National Science Foundation.}  After
masking out the H-$\gamma$ and H-$\delta$ lines, we cross correlated
in the wavelength range 400--460 nm between our observed spectra and a
theoretically generated template spectrum from \citet{coe05} interpolated
to $T_{\mathrm{eff}} = 5440$, $\log{g} = 3.0$, and $\mathrm{[Fe/H]} =
-2.45$.  After applying heliocentric corrections, we give our final radial
velocity measurements as well as their uncertainties and the Heliocentric
Julian Date (HJD) at the start of each observation in Table~\ref{tbl-2}.

\begin{deluxetable}{LRR}
\tablecaption{High-resolution Radial Velocity Observations of 2MASS
J18082002-5104378 A\label{tbl-2}}
\tablewidth{0pt}
\tablehead{
\colhead{HJD} & \colhead{Radial Velocity} & \colhead{Uncertainty} \\
\colhead{(d)} & \colhead{(km s$^{-1}$)} & \colhead{(km s$^{-1}$)}}
\startdata
\textrm{VLT/UVES} & & \\
2456949.52132 & 21.19 & 0.26 \\
2456951.51539 & 18.15 & 0.24 \\
2457087.86838 & 21.94 & 0.26 \\
\hline
\textrm{Magellan/MIKE} & & \\
2457479.88381 &  8.76 & 0.62 \\
2457480.90651 &  7.47 & 0.40 \\
2457579.66548 & 13.57 & 0.54 \\
2457581.73107 & 10.63 & 0.32 \\
2457582.64291 & 10.29 & 0.50 \\
2457590.69448 &  9.61 & 0.32 \\
2457591.59471 & 10.63 & 0.36 \\
2457593.53991 & 13.48 & 0.40 \\
2457595.46615 & 16.71 & 0.92 \\
2457680.48650 & 19.19 & 0.66 \\
2457946.72530 & 21.94 & 0.28 \\
2457948.57123 & 24.40 & 0.28 \\
2457949.59345 & 25.12 & 0.26 \\
2457953.78650 & 25.55 & 0.32 \\
\enddata
\end{deluxetable}

We found it impossible to fit a Keplerian orbit to the union of our
Magellan/MIKE radial velocities given in Table~\ref{tbl-2} and the
Very Large Telescope (VLT) Ultraviolet and Visual Echelle Spectrograph
\citep[UVES;][]{dek00} radial velocities published in \citet{mel16}.
After measuring the radial velocities ourselves for the VLT/UVES spectra
described in \citet{mel16}, we found that those authors applied the
heliocentric correction to their measured velocities with the wrong sign.
Instead of the presence of a neutron star or stellar mass black hole in
the system, it was that mistake that produced the apparent 100 km s$^{-1}$
change in the radial velocity of the 2MASS J18082002-5104378 system over
a five month interval.  We report our own radial velocity measurements
based on the \citet{mel16} VLT/UVES spectra in the first three rows of
Table~\ref{tbl-2}.  It is clear from Table~\ref{tbl-2} that the 2MASS
J18082002-5104378 system is a single-lined spectroscopic binary, and we
describe our Keplerian fit to those data in the following subsection.

We used a two-step procedure to measure radial velocities from our
Gemini South/GMOS-S spectra.  We first cross-correlated each exposure
against a rest-frame continuum-normalized spectrum of the UMP giant
CD--38 245.  We then corrected each exposure for the measured radial
velocity shift and produced a continuum-normalized spectrum of 2MASS
J18082002-5104378 using the median flux in each rebinned exposure.
We next used the stacked rest-frame spectrum of 2MASS J18082002-5104378
as the template and re-measured the radial velocities of individual
exposures by cross-correlation.  The radial velocities we report in
Table~\ref{tbl-3} represent the mean heliocentric-corrected radial
velocity from three exposures.  The listed uncertainty is the standard
deviation of three measurements added in quadrature with a systematic
uncertainty equal to the mean estimated uncertainty of individual
measurements from cross-correlation (about 5 km s$^{-1}$).

While less precise than the data in Table~\ref{tbl-2}, the data in
Table~\ref{tbl-3} have better time resolution and sufficient radial
velocity precision to recover any large-amplitude radial velocity
variations if present.  We did not observe any large-amplitude radial
velocity variations over a 13 month period, supporting our analysis
in the preceding paragraph.  We experimented with fitting the data in
Table~\ref{tbl-2} alone and in concert with the data in Table~\ref{tbl-3}.
We found that the resulting Keplerian orbital elements were consistent
between the two cases but less precise in the latter case.  Consequently,
we did not use the data in Table~\ref{tbl-3} in our final Keplerian fit
described in the next subsection.

\begin{deluxetable}{LRR}
\tablecaption{Gemini South/GMOS-S Low-resolution Radial Velocity
Observations of 2MASS J18082002-5104378 A\label{tbl-3}}
\tablewidth{0pt}
\tablehead{
\colhead{HJD} & \colhead{Radial Velocity} & \colhead{Uncertainty} \\
\colhead{(d)} & \colhead{(km s$^{-1}$)} & \colhead{(km s$^{-1}$)}}
\startdata
2457559.08034 &  8.5 & 5.7 \\
2457563.32367 & 20.9 & 5.3 \\
2457584.09253 & -6.0 & 5.7 \\
2457590.06602 &  6.3 & 5.5 \\
2457603.99236 & 26.3 & 6.2 \\
2457634.97860 & 23.0 & 5.1 \\
2457654.00692 &  7.4 & 5.1 \\
2457663.99389 & 15.7 & 5.1 \\
2457681.01178 & 21.1 & 5.6 \\
2457683.07641 &  5.2 & 5.6 \\
2457686.02310 & 14.9 & 5.1 \\
2457808.38897 & 26.4 & 5.8 \\
2457815.36205 & 21.0 & 5.2 \\
2457836.37176 & 13.4 & 5.1 \\
2457842.32615 & 23.9 & 5.2 \\
2457850.41464 & 32.2 & 5.2 \\
2457869.34934 & 10.0 & 5.0 \\
2457872.35197 & 21.2 & 5.1 \\
2457883.17645 & 22.3 & 5.2 \\
2457890.38603 & 16.0 & 5.7 \\
2457901.30690 &  4.6 & 5.8 \\
2457923.23911 & 28.2 & 5.2 \\
2457934.12196 &  9.0 & 5.3 \\
2457944.10230 & 20.2 & 5.3 \\
2457948.03707 & 28.7 & 6.3 \\
2457954.20577 & 20.9 & 6.6 \\
2457955.20855 & 31.1 & 5.3 \\
2457956.08313 & 23.1 & 5.3 \\
2457957.31008 & 18.0 & 5.3 \\
2457958.06814 & 19.8 & 5.3 \\
2457961.03876 & 19.4 & 5.6 \\
\enddata
\end{deluxetable}

\subsection{Keplerian Orbit Parameter Estimation}

\begin{figure*}
\plotone{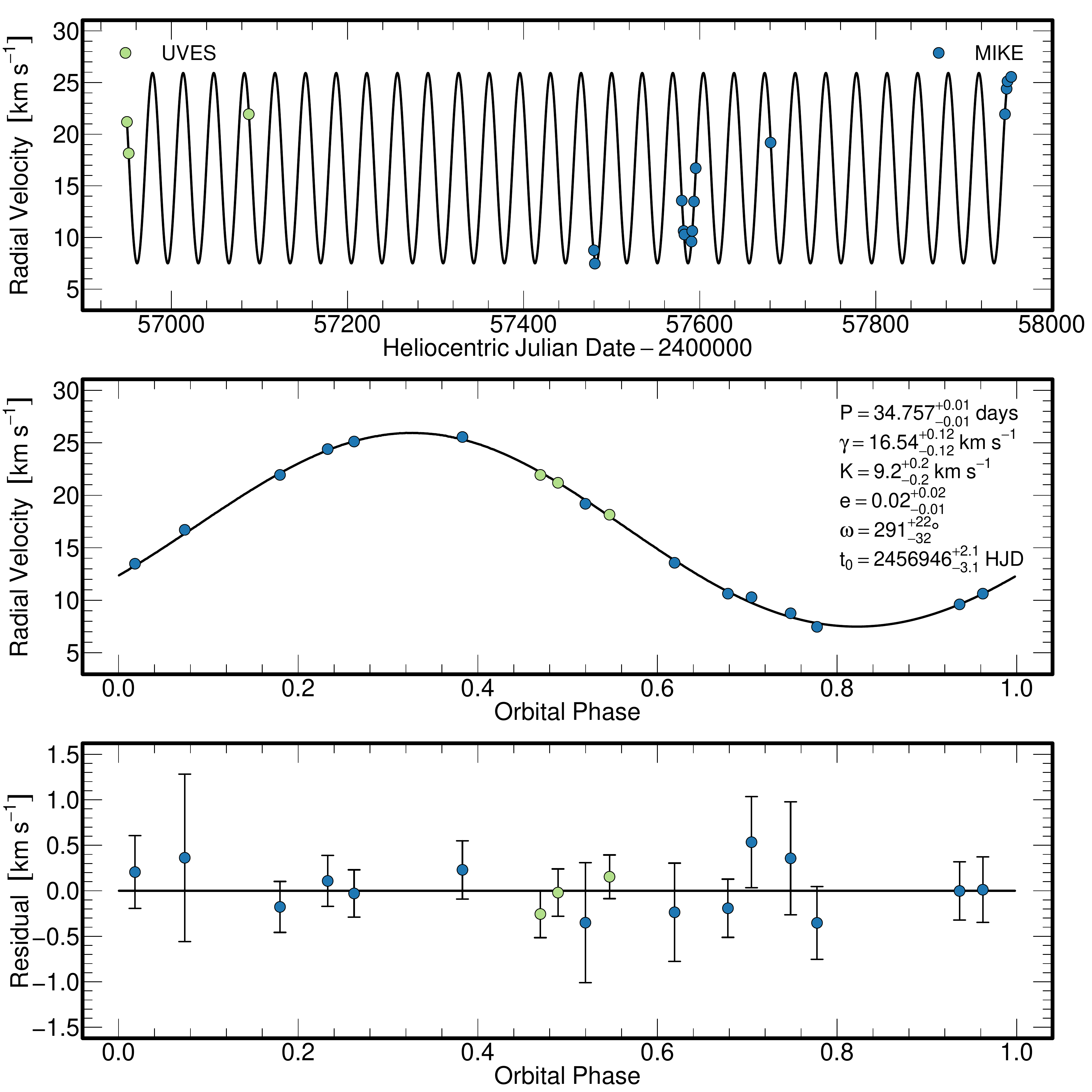}
\caption{Keplerian orbit of 2MASS J18082002--5104378 system alongside
radial velocities measured with VLT/UVES and Magellan/MIKE.  The points
are the individual radial velocities from Table~\ref{tbl-2}, while
the curve is the orbit resulting from a Keplerian fit to those data.
We plot the radial velocity uncertainties in each panel, but they are
smaller than the plotted points in the top two panels.\label{fig01}}
\end{figure*}

\begin{figure*}
\plotone{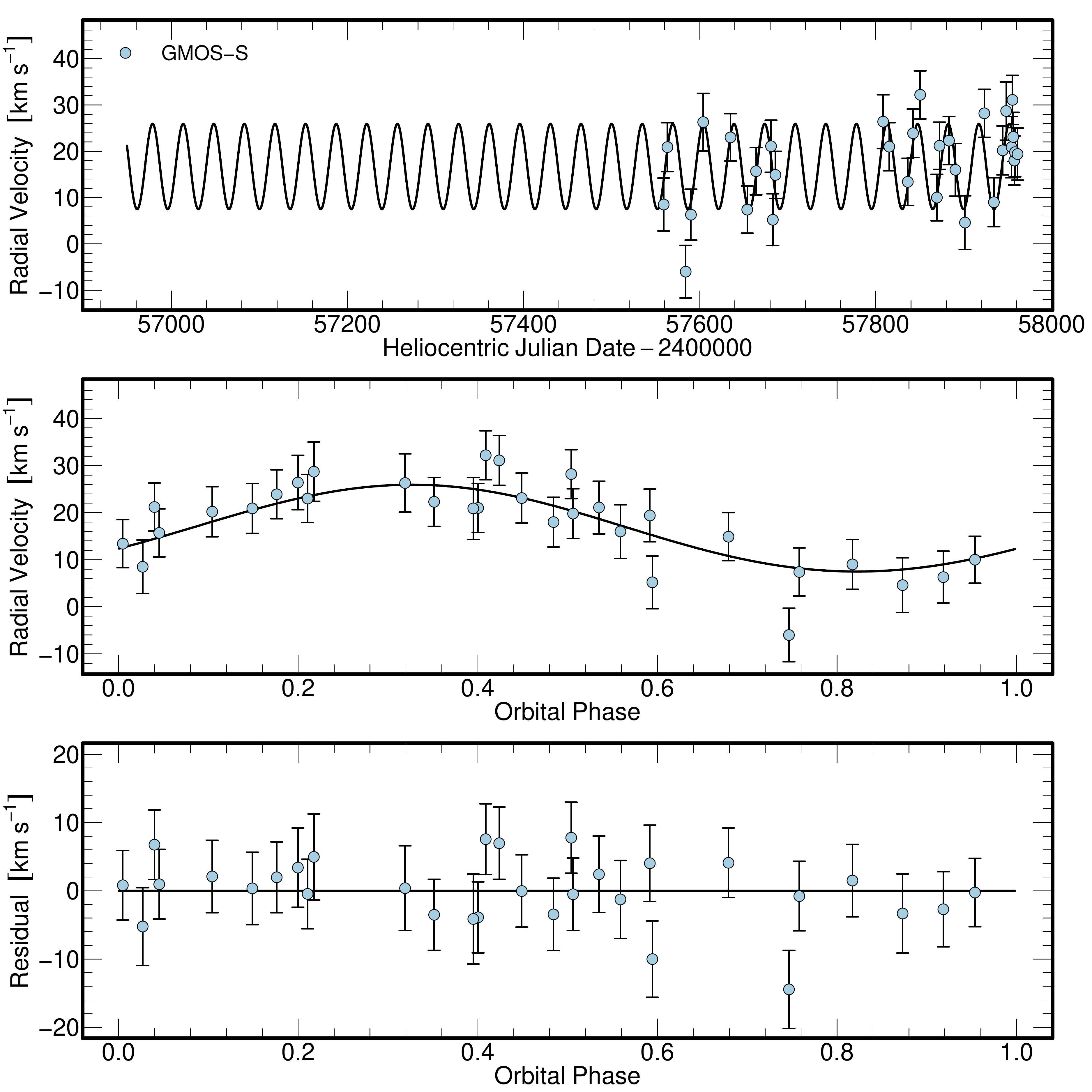}
\caption{Keplerian orbit of 2MASS J18082002--5104378 system alongside
radial velocities measured with Gemini South/GMOS-S.  The points
are the individual radial velocities from Table~\ref{tbl-3}, while
the curve is the orbit resulting from a Keplerian fit to data in
Table~\ref{tbl-2}.\label{fig02}}
\end{figure*}

\begin{figure*}
\plotone{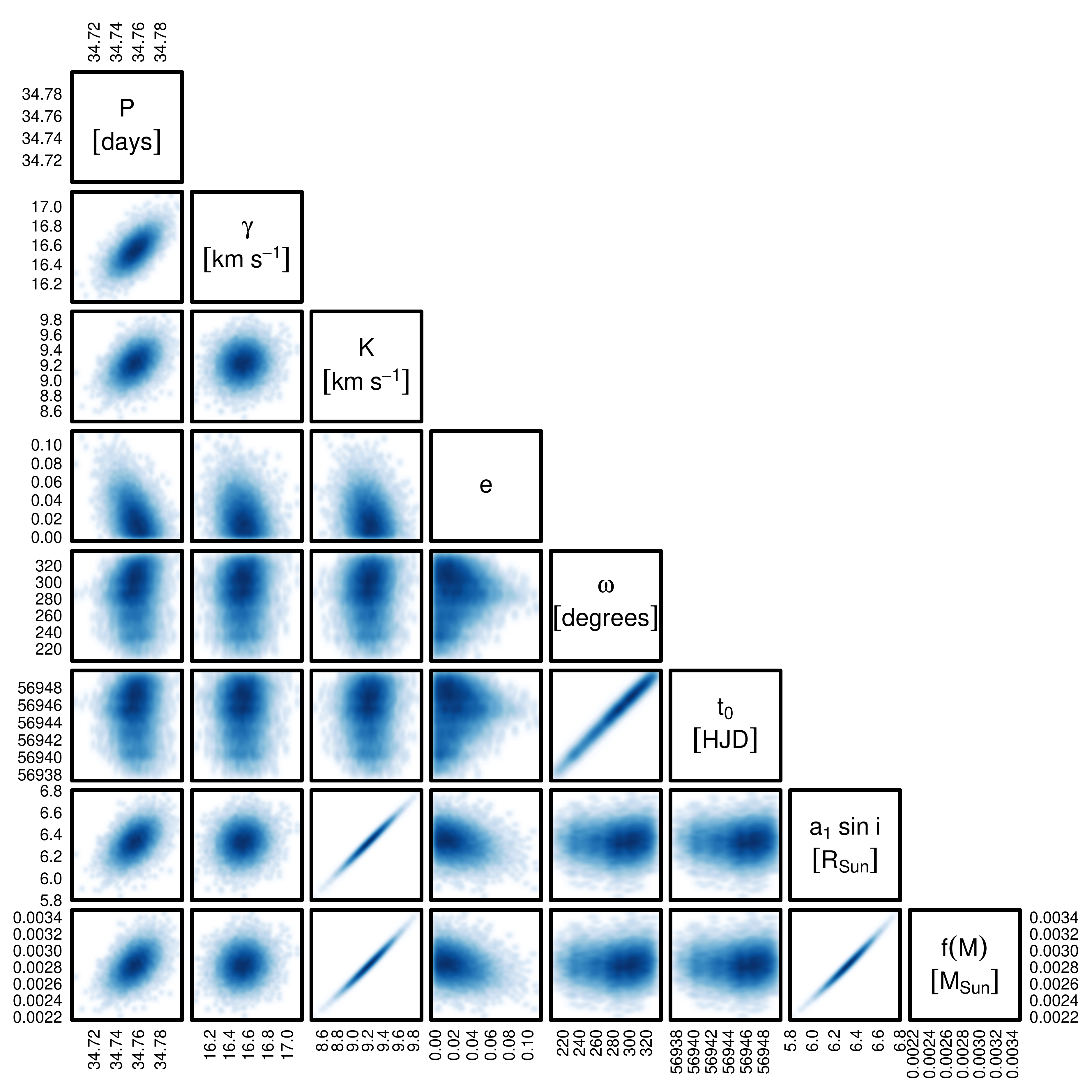}
\caption{Corner plot resulting from our MCMC fit of a Keplerian
orbit to the radial velocities of the 2MASS J18082002--5104378
system given in Table~\ref{tbl-2}.  We found an orbital period $P =
34.757 \pm 0.010$ days, a system velocity $\gamma = 16.54 \pm 0.12$
km s$^{-1}$, a velocity semi-amplitude $K = 9.2 \pm 0.2$ km s$^{-1}$,
an eccentricity $e = 0.02_{-0.01}^{+0.02}$, a longitude of periastron
$\omega = 291_{-32}^{+22}$ deg, and a time of periastron $t_{0} =
2456945.9_{-3.1}^{+2.1}$ HJD.\label{fig03}}
\end{figure*}

We fit a Keplerian orbit to the radial velocities in
Table~\ref{tbl-2} using the Bayesian Markov chain Monte Carlo (MCMC) code
\texttt{ExoFit}\footnote{\url{http://www.homepages.ucl.ac.uk/~ucapola/exofit.html}}
\citep{bal09}.  We used the default \texttt{ExoFit} parameters with one
exception: we initialized the starting point of the MCMC at $e = 0.1$
instead of $e = 0.5$.  We plot the fitted Keplerian orbit with the radial
velocities from Tables~\ref{tbl-2} and~\ref{tbl-3} in Figures~\ref{fig01}
and~\ref{fig02}.  We include a corner plot illustrating the covariances
of the fitted parameters in Figure~\ref{fig03}.  We found an orbital
period $P = 34.757 \pm 0.010$ days, a system velocity $\gamma = 16.54
\pm 0.12$ km s$^{-1}$, a velocity semi-amplitude $K = 9.2 \pm 0.2$
km s$^{-1}$, an eccentricity $e = 0.02_{-0.01}^{+0.02}$, a longitude
of periastron $\omega = 291_{-32}^{+22}$ deg, and a time of periastron
$t_{0} = 2456945.9_{-3.1}^{+2.1}$ HJD.  We also calculated the projected
semimajor axis $a_{1} \sin{i}$ as

\begin{eqnarray}
a_{1} \sin{i} & = & \frac{P K \left(1-e^{2}\right)^{1/2}}{2 \pi} \\
              & = & 6.3 \pm 0.1~R_{\odot}, \nonumber
\end{eqnarray}

\noindent
and the mass function $f(M)$ as

\begin{eqnarray}
f(M) & = & \frac{P K^{3} \left(1-e^{2}\right)^{3/2}}{2 \pi G} \\
     & = & 0.0028 \pm 0.0001~M_{\odot}. \nonumber
\end{eqnarray}

\noindent
We summarize the properties of the 2MASS J18082002--5104378 system
in Table~\ref{tbl-4}.\footnote{Though we did not use the available
\textit{Gaia} DR2 \citep{sal17,eva18,rie18} or APASS photometry
\citep{hen16} for 2MASS J18082002--5104378, we provide those data in
Table~\ref{tbl-4} for context.}

\begin{deluxetable*}{lcl}
\tablecaption{2MASS J18082002--5104378 System Properties\label{tbl-4}}
\tablewidth{0pt}
\tablehead{
\colhead{Property} & \colhead{Value} & \colhead{Units}}
\startdata
Observed Properties & & \\
Gaia DR2 R.A. $\alpha$ (J2000) & 18 08 20.0314 & h m s \\
Gaia DR2 decl. $\delta$ (J2000) & -51 04 37.884 & d m s \\
Gaia DR2 galactic longitude $l$ (J2000) & 342.5353 & degrees \\
Gaia DR2 galactic latitude $b$ (J2000) & -14.4898 & degrees \\
Gaia DR2 proper motion $\mu_{\alpha} \cos{\delta}$ & $-5.63 \pm 0.07$ & mas yr$^{-1}$ \\
Gaia DR2 proper motion $\mu_{\delta}$ & $-12.64 \pm 0.06$ & mas yr$^{-1}$ \\
Gaia DR2 parallax $\pi$ & $1.64 \pm 0.04$ & mas \\
Gaia DR2 $G$ & $11.7562 \pm 0.0003$ & Vega mag \\
Gaia DR2 $G_{\mathrm{BP}}$ & $12.119 \pm 0.002$ & Vega mag \\
Gaia DR2 $G_{\mathrm{RP}}$ & $11.215 \pm 0.002$ & Vega mag \\
APASS $B$ & $12.546 \pm 0.018$ & Vega mag \\
APASS $V$ & $11.930 \pm 0.031$ & Vega mag \\
APASS $g'$ & $12.219 \pm 0.014$ & AB mag \\
APASS $r'$ & $11.736 \pm 0.019$ & AB mag \\
APASS $i'$ & $11.625 \pm 0.017$ & AB mag \\
SkyMapper $u$ & $13.331 \pm 0.005$ & AB mag \\
SkyMapper $v$ & $12.892 \pm 0.002$ & AB mag \\
SkyMapper $g$ & $12.067 \pm 0.002$ & AB mag \\
SkyMapper $r$ & $11.747 \pm 0.002$ & AB mag \\
2MASS $J$ & $10.527 \pm 0.026$ & Vega mag \\
2MASS $H$ & $10.158 \pm 0.026$ & Vega mag \\
2MASS $K_{\mathrm{s}}$ & $10.088 \pm 0.026$ & Vega mag \\
\textit{WISE} $W1$ & $9.992 \pm 0.023$ & Vega mag \\
\textit{WISE} $W2$ & $9.997 \pm 0.020$ & Vega mag \\
\textit{WISE} $W3$ & $9.919 \pm 0.054$ & Vega mag \\
Orbital period $P$ & $34.757 \pm 0.010$ & day \\
System velocity $\gamma$ & $16.54 \pm 0.12$ & km s$^{-1}$ \\
Velocity semi-amplitude $K$ & $9.2 \pm 0.2$ & km s$^{-1}$ \\
Eccentricity $e$ & $0.02_{-0.01}^{+0.02}$ & \\
Longitude of periastron $\omega$ & $291_{-32}^{+22}$ & deg \\
Time of periastron $t_{0}$ & $2456945.9_{-3.1}^{+2.1}$ & HJD \\
Projected semimajor axis $a_{1} \sin{i}$ & $6.3 \pm 0.1$ & $R_{\odot}$ \\
Mass function $f(M)$ & $0.0028 \pm 0.0001$ & $M_{\odot}$ \\
Time span & $28.894 \pm 0.008$ & periods \\
\hline
Inferred Properties & & \\
Primary mass $M_{1}$ & $0.7599 \pm 0.0001$ & $M_{\odot}$ \\
System age $\tau$ & $13.53 \pm 0.002$ & Gyr \\
Minimum secondary mass $M_{2,\mathrm{min}}$ & $0.131 \pm 0.002$ & $M_{\odot}$ \\
Secondary mass $M_{2}$ & $0.14_{-0.01}^{+0.06}$ & $M_{\odot}$ \\
Semimajor axis $a$ & $0.202_{-0.001}^{+0.004}$ & au \\
Total Galactic velocity $v$ & $207.5_{-1.2}^{+1.1}$ & km s$^{-1}$ \\
Pericenter of Galactic orbit $R_{\mathrm{peri}}$ & $5.56 \pm 0.07$ & kpc \\
Apocenter of Galactic orbit $R_{\mathrm{apo}}$ & $7.66 \pm 0.02$ & kpc \\
Eccentricity of Galactic orbit $e_{\mathrm{G}}$ & $0.158_{-0.004}^{+0.005}$ & \\
Maximum distance from Galactic plane $z_{\mathrm{max}}$ & $0.126_{-0.003}^{+0.005}$ & kpc \\
\enddata
\end{deluxetable*}

The mass function given above indicates that the visible
star is the primary in the system which we now denote 2MASS
J18082002--5104378 A.  To estimate the properties of the
unseen component 2MASS J18082002--5104378 B, we first used the
\texttt{isochrones}\footnote{\url{https://github.com/timothydmorton/isochrones}}
\citep{mor15} package to estimate the mass of the visible star 2MASS
J18082002--5104378 A using as inputs its:

\begin{enumerate}
\item
estimated spectroscopic parameters and associated uncertainties from
\citet{mel16};
\item
$u$, $v$, $g$, and $r$ magnitudes and associated uncertainties from Data
Release (DR) 1.1 of the SkyMapper Southern Sky Survey \citep{wol18};
\item
$J$, $H$, and $K_{s}$ magnitudes and associated uncertainties from the
2MASS All-sky Point Source Catalog \citep{skr06};
\item
$W1$, $W2$, and $W3$ magnitudes and associated uncertainties from the
AllWISE Source Catalog \citep{wri10,mai11};
\item
\textit{Gaia} DR2 parallax and uncertainty
\citep{gai16,gai18,are18,ham18,lin18,lur18}.
\end{enumerate}

\noindent
We used \texttt{isochrones} to fit the Dartmouth Stellar
Evolution Database \citep{dot07,dot08} library generated with the
Dartmouth Stellar Evolution Program (DSEP) to these observables using
\texttt{MultiNest}\footnote{\url{https://ccpforge.cse.rl.ac.uk/gf/project/multinest/}}
\citep{fer08,fer09,fer13}.  We restricted the Dartmouth library to
$\alpha$-enhanced composition $[\alpha\mathrm{/Fe}] = +0.4$, stellar age
$\tau$ in the range 10.0 Gyr $\leq \tau \leq$ 13.721 Gyr, and extinction
$A_{V}$ in the range 0 mag $\leq A_{V} \leq$ 1.0 mag.  We limited
distances $d$ considered to the range 561.7978 pc $\leq d \leq$ 632.9114
pc (the 2-$\sigma$ range for the system from \textit{Gaia} DR2).  Given
these constraints, we found that the primary 2MASS J18082002--5104378
A has a mass $M_{1} =  0.7599 \pm 0.0001~M_{\odot}$ and an age $\tau =
13.535 \pm 0.002$ Gyr.  We summarize our preferred isochrone-derived
parameters for 2MASS J18082002--5104378 A in Table~\ref{tbl-5}.

The Dartmouth isochrones prefer a slightly warmer and more metal-rich star
than suggested by \citet{mel16}.  The estimated effective temperature
is degenerate with extinction, however, and that degeneracy may explain
the slightly higher metallicity suggested by the isochrone analysis.
It is also possible that 2MASS J18082002--5104378 A is slightly carbon
enhanced, as the \citet{mel16} upper limit on the carbon abundance
[C/Fe] is $[\mathrm{C/Fe}] \lesssim +0.5$.  If 2MASS J18082002--5104378
A has $[\mathrm{C/Fe}] \approx +0.5$, then that would increase its
total metallicity $[\mathrm{M/H}]$ to $[\mathrm{M/H}] \approx -3.50$
in accord with the isochrone estimate.

\begin{deluxetable}{lcl}
\tablecaption{Isochrone-derived Stellar Parameters for 2MASS
J18082002--5104378 A\label{tbl-5}}
\tablewidth{0pt}
\tablehead{
\colhead{Property} & \colhead{Value} & \colhead{Units}}
\startdata
Effective temperature $T_{\mathrm{eff}}$ & $5871_{-18}^{+17}$ & K \\
Surface gravity $\log{g}$ & $3.378_{-0.006}^{+0.007}$ & \\
Metallicity [Fe/H] & $-3.50 \pm 0.02$ & \\
Extinction $A_{V}$ & $0.40 \pm 0.01$ & mag \\
Mass $M_{\ast}$ & $0.7599 \pm 0.0001$ & $M_{\odot}$ \\
System age $\tau$ & $13.535 \pm 0.002$ & Gyr \\
\enddata
\end{deluxetable}

To investigate the possibility of systematic error resulting from
our use of the Dartmouth library, we also estimated the mass and
age of 2MASS J18082002--5104378 A using two additional isochrone
libraries.  We first used \texttt{isochrones} to fit the MESA
Isochrones \& Stellar Tracks \citep[MIST;][]{cho16,dot16} library
generated with the Modules for Experiments in Stellar Astrophysics
\citep[MESA;][]{pax11,pax13,pax15} code to these observables using
\texttt{MultiNest}.  We found $M_{1} = 0.80728_{-0.00005}^{+0.00004}
~M_{\odot}$ and $\tau = 11.12 \pm 0.07$ Gyr.  We then used the
PARAM 1.3 web interface for the Bayesian estimation of stellar
parameters\footnote{\url{http://stev.oapd.inaf.it/cgi-bin/param_1.3}}
\citep{das06}.  We found $M_{1} = 0.768 \pm 0.011~M_{\odot}$ and $\tau
= 12.747 \pm 0.553$ Gyr.  We prefer the mass and age produced by the
Dartmouth grid because it accounts for the $\alpha$-enhanced composition
of 2MASS J18082002--5104378 A and because its estimate has a higher
log-likelihood than either the MIST or PARAM 1.3 estimates.

To estimate the mass of 2MASS J18082002--5104378 B, we solved the
nonlinear mass function equation

\begin{eqnarray}
\frac{M_{1} \sin^{3}\!{i}}{f(m)} q^{3} - (1+q)^{2} = 0,
\end{eqnarray}

\noindent
where $q = M_{2}/M_{1}$ and the mass function $f(m)$ is defined as

\begin{eqnarray}
f(m) = \frac{M_{2}^{3} \sin^{3}\!{i}}{\left(M_{1} + M_{2}\right)^{2}}
     = \frac{P K^{3} \left(1-e^{2}\right)^{3/2}}{2 \pi G}.
\end{eqnarray}

We used the distribution of $\sin^{3}\!{i}$ proposed by \citet{hog92}.
Generalizing the method put forward by \citet{hal87}, Hogeveen showed that
the assumption $P(i)\,di = (4/\pi) \sin^{2}\!{i}\,di$ produces better
estimates of the measured mass ratios of double-lined spectroscopic
binaries than the standard assumption of random inclinations $P(i)\,di =
\sin{i}\,di$.  This bias toward edge-on systems arises because $K$-limited
searches for spectroscopic binaries will tend to discover systems with
orbital inclinations closer to $i = 90^{\circ}$ than to $i = 0^{\circ}$.
Under the \citet{hog92} assumption, the distribution of $x= \sin^{3}\!{i}$
becomes

\begin{eqnarray}
P(x)\,dx = \frac{4}{3 \pi} \left(1 - x^{2/3}\right)^{-1/2}\,dx.
\end{eqnarray}

\begin{figure*}
\plotone{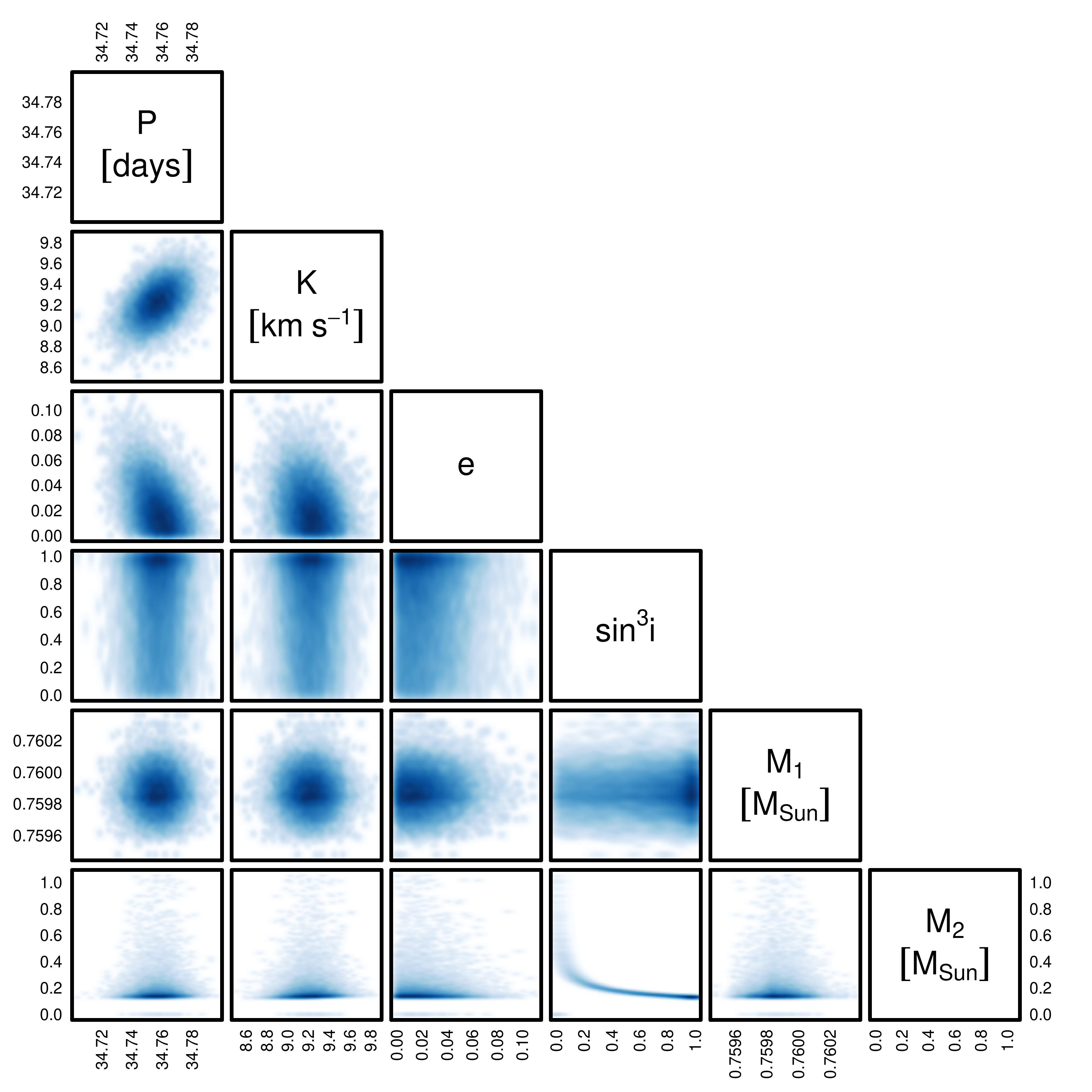}
\caption{Corner plot for our estimate of the mass of the secondary
in the 2MASS J18082002--5104378 system.  The posteriors on $P$, $K$,
and $e$ came from our MCMC fit of a Keplerian orbit to the radial
velocities in Table~\ref{tbl-2}.  The distribution of $\sin^{3}\!{i}$
is from \citet{hog92}.  The posterior on the mass of the primary
$M_{1}$ came from our isochrone analysis.  The mass of the secondary
$M_{2}$ came from solving the nonlinear mass function equation.
We found that the mass of the secondary 2MASS J18082002--5104378 B is
$M_{2} = 0.14_{-0.01}^{+0.06}~M_{\odot}$, making it the lowest-mass
UMP star known and placing it near the hydrogen-burning limit at
$M_{\ast} \approx 0.092~M_{\odot}$ for a star of its composition
\citep{sau94}.\label{fig04}}
\end{figure*}

We found a mass $M_{2} = 0.14_{-0.01}^{+0.06}~M_{\odot}$ and a minimum
mass $M_{2,\mathrm{min}} = 0.131 \pm 0.002~M_{\odot}$ (assuming
$\sin^{3}\!{i} = 1$).  If we instead assume a single average value
for $<\!\sin^{3}\!{i}\!>$ = 0.679 in place of averaging over the
entire distribution of $\sin^{3}\!{i}$, we find $M_{2} = 0.151 \pm
0.003~M_{\odot}$.  This inferred mass makes 2MASS J18082002--5104378 B
the lowest-mass UMP star known.  Indeed, it is near the hydrogen-burning
limit at $M_{\ast} \approx 0.092~M_{\odot}$ for a star of its composition
\citep{sau94}.  We provide a corner plot illustrating the covariances of
the parameters necessary for the mass calculation in Figure~\ref{fig04}.

We calculated an upper limit on the mass of 2MASS J18082002--5104378 B
based on the lack of a second peak in our cross-correlation analysis.
The UVES data we analyzed have $\mathrm{S/N} \approx 50$ pixel$^{-1}$
between 400 and 460 nm, so a secondary star with a similar spectrum to
2MASS J18082002--5104378 A that is less than three magnitudes fainter
in the $B$ band would be detectable.  Using a MIST isochrone with the
same age we calculated for 2MASS J18082002--5104378 A, that corresponds
to an upper limit on the secondary mass $M_{2,\mathrm{max}} \approx
0.6~M_{\odot}$.

\subsection{Galactic Orbit Estimation}

To better understand the origin of the 2MASS J18082002--5104378
system, we calculated its Galactic orbit using
\texttt{galpy}\footnote{\url{https://github.com/jobovy/galpy}}.
We sampled 1000 Monte Carlo realizations from the \textit{Gaia} DR2
astrometric solution for 2MASS J18082002--5104378 taking full account of
the covariances between position, parallax, and proper motion.  We used
our posterior on the system radial velocity $\gamma$ and assumed no
covariance between our measured radial velocity and the \textit{Gaia}
DR2 astrometric solution.  We used each Monte Carlo realization as
an initial condition for an orbit and integrated it forward 10 Gyr in
a Milky Way-like potential.  We adopted the \texttt{MWPotential2014}
described by \citet{bov15}.  In that model, the bulge is parameterized
as a power-law density profile that is exponentially cut-off at 1.9
kpc with a power-law exponent of $-1.8$.  The disk is represented by
a Miyamoto--Nagai potential with a radial scale length of 3 kpc and a
vertical scale height of 280 pc \citep{miy75}.  The halo is modeled as a
Navarro--Frenk--White halo with a scale length of 16 kpc \citep{nav96}.
We set the solar distance to the Galactic center to $R_{0} = 8.122$, kpc,
the circular velocity at the Sun to $V_{0} = 238$ km s$^{-1}$, the height
of the Sun above the plane to $z_{0} = 25$ pc, and the solar motion with
the respect to the local standard of rest to ($U_{\odot}$, $V_{\odot}$,
$W_{\odot}$) = (10.0, 11.0, 7.0) km s$^{-1}$ \citep{jur08,bla16,gra18}.

We found that the 2MASS J18082002--5104378 system is on a thin disk
like orbit through the Galaxy with pericenter $R_{\mathrm{peri}} =
5.56 \pm 0.07$ kpc, apocenter $R_{\mathrm{apo}} = 7.66 \pm 0.02$ kpc,
Galactic eccentricity $e_{\mathrm{G}} = 0.158_{-0.004}^{+0.005}$,
and maximum distance from Galactic plane $z_{\mathrm{max}}
= 0.126_{-0.003}^{+0.005}$ kpc.  By some margin, the 2MASS
J18082002--5104378 system is the most metal-poor star system on a thin
disk orbit \citep[e.g.,][]{cas11,ben14,bee17}.

\subsection{Importance for Pop III Star Formation}

To evaluate the implications for Pop III stars of our discovery of the
low-mass UMP star 2MASS J18082002--5104378 B, we explored the properties
and survival of fragments in models of UMP and primordial composition
protostellar disks published in \citet{tan14}.  A Keplerian disk can be
expected to fragment if its Toomre parameter $Q$ is

\begin{eqnarray}
Q = \frac{\Omega c_{s}}{\pi G \Sigma} \lesssim 1,
\end{eqnarray}

\noindent
where $\Omega$ is the Keplerian orbital frequency, $c_{s}$ is the gas
sound speed, and $\Sigma$ is the gas surface density.  $\Omega$, $c_{s}$,
and $\Sigma$ are all implicit functions of the radial coordinate $r$.
In this situation, the typical fragment mass scale is

\begin{eqnarray}\label{mfrag}
M_{\mathrm{frag}} \sim \Sigma H^{2},
\end{eqnarray}

\noindent
where $H = c_{s}/\Omega$ is the characteristic disk thickness
\citep[e.g.,][]{paa18}.  A scaling relation for the mass of the typical
fragment that will form in an unstable disk is therefore

\begin{eqnarray}\label{scaling1}
\frac{M_{\mathrm{frag},2}}{M_{\mathrm{frag,1}}} = \left(\frac{\Sigma_{2}}{\Sigma_{1}}\right) \left(\frac{H_{2}}{H_{1}}\right)^2.
\end{eqnarray}

Once formed, a fragment in a nonaxisymmetric protostellar disk will
lose orbital energy and angular momentum due to gravitational torques
between the fragment and the disk.  As shown by \citet{bar11}, fragments
formed via gravitational instability will not be massive enough relative
to the local disk mass to significantly perturb the disk structure.
Assuming the fragment moves on a circular orbit at the local Keplerian
frequency, the characteristic fragment migration time is given by

\begin{eqnarray}\label{tmig}
t_{\mathrm{mig}} = \frac{h^2}{q} \frac{M_{\ast}}{r^2 \Sigma} \Omega^{-1},
\end{eqnarray}

\noindent
where $h = H/r$ is the disk aspect ratio and $q =
M_{\mathrm{frag}}/M_{\ast}$ is ratio between the fragment mass and stellar
mass \citep[e.g.,][]{paa18}.  A scaling relation for the typical fragment
migration time in an unstable disk is therefore

\begin{eqnarray}\label{scaling2}
\frac{t_{\mathrm{mig},2}}{t_{\mathrm{mig},1}} & = & \left(\frac{h_{2}}{h_{1}}\right)^2 \left(\frac{q_{1}}{q_{2}}\right) \left(\frac{M_{\ast,2}}{M_{\ast,1}}\right) \nonumber \\
& & \left(\frac{r_{1}}{r_{2}}\right)^2 \left(\frac{\Sigma_{1}}{\Sigma_{2}}\right) \left(\frac{\Omega_{1}}{\Omega_{2}}\right).
\end{eqnarray}

To calculate the disk properties necessary to use the scaling relations
given by Equations (\ref{scaling1}) and (\ref{scaling2}), we used the
\citet{tan14} models and assumed a mean molecular weight $\mu = 2.29$
for the fully molecular gas expected at protostellar disk densities
\citep[e.g.,][]{sch12}.  We report all values necessary to use Equations
(\ref{scaling1}) and (\ref{scaling2}) in Table~\ref{tbl-6}.

\begin{deluxetable*}{lcccccccc}
\tablecaption{Protostellar Disk Properties from \citet{tan14}\label{tbl-6}}
\tablewidth{0pt}
\tablehead{
\colhead{Composition} & \colhead{$M_{\ast}$}     & \colhead{$r$}  & \colhead{$T$} & \colhead{$\rho$}        & \colhead{$c_{s}$}       &  \colhead{$\Omega$}  & \colhead{$H$}  & \colhead{$\Sigma$} \\
                      & \colhead{($M_{\odot}$)}  & \colhead{(au)} & \colhead{(K)} & \colhead{(g cm$^{-3}$)} & \colhead{(cm s$^{-1}$)} & \colhead{(s$^{-1}$)} & \colhead{(au)} & \colhead{(g cm$^{-2}$)}}
\startdata
$Z = 10^{-4} Z_{\odot}$ &   1 & $10^{0.5}$ & $5.0 \times 10^{2}$ & $1.7 \times 10^{-9}$  & $1.3 \times 10^{5}$ & $3.5 \times 10^{-8}$ & 0.25 & $6.3 \times 10^{3}$ \\
                        &   1 & $10^{1.0}$ & $2.0 \times 10^{2}$ & $8.4 \times 10^{-11}$ & $8.5 \times 10^{4}$ & $6.3 \times 10^{-9}$ & 0.90 & $1.1 \times 10^{3}$ \\
                        &   1 & $10^{1.5}$\tablenotemark{a} & $1.0 \times 10^{2}$ & $6.7 \times 10^{-12}$ & $6.0 \times 10^{4}$ & $1.1 \times 10^{-9}$ &  3.6 & $3.6 \times 10^{2}$ \\
$Z = 0$                 &  10 & $10^{0.5}$ & $1.6 \times 10^{3}$ & $1.7 \times 10^{-8}$  & $2.4 \times 10^{5}$ & $1.1 \times 10^{-7}$ & 0.14 & $3.6 \times 10^{4}$ \\
                        &  10 & $10^{1.0}$ & $2.0 \times 10^{3}$ & $5.3 \times 10^{-10}$ & $2.7 \times 10^{5}$ & $2.0 \times 10^{-8}$ & 0.90 & $7.1 \times 10^{3}$ \\
                        &  10 & $10^{2.1}$\tablenotemark{a} & $7.9 \times 10^{2}$ & $3.3 \times 10^{-13}$ & $1.7 \times 10^{5}$ & $4.5 \times 10^{-10}$ &  25 & $1.3 \times 10^{2}$ \\
$Z = 0$                 & 100 & $10^{0.5}$ & $4.0 \times 10^{4}$ & $1.7 \times 10^{-8}$  & $1.2 \times 10^{6}$ & $3.5 \times 10^{-7}$ & 0.23 & $5.7 \times 10^{4}$ \\
                        & 100 & $10^{1.0}$ & $1.6 \times 10^{3}$ & $6.7 \times 10^{-9}$  & $2.4 \times 10^{5}$ & $6.3 \times 10^{-8}$ & 0.25 & $2.5 \times 10^{4}$ \\
                        & 100 & $10^{3.2}$\tablenotemark{a} & $3.2 \times 10^{2}$ & $4.2 \times 10^{-15}$ & $1.1 \times 10^{5}$ & $3.2 \times 10^{-11}$ & 230 & $1.4 \times 10^{1}$ \\
\enddata
\tablenotetext{a}{Outer edge of disk}
\end{deluxetable*}

Because it is so tightly bound to J18082002--5104378 A, 2MASS
J18082002--5104378 B is extraordinarily unlikely to have been
gravitationally captured outside of the birth environment of the
former.  They must therefore have formed in the same molecular core
and thus have the same composition.  It is reasonable to assume that
the short-period, low-mass star 2MASS J18082002--5104378 B formed via
disk fragmentation and arrived close to its host star because of disk
migration \citep[e.g.,][]{kra16}.  In that case, its formation and
survival at $M_{\ast} \approx 0.14~M_{\odot}$ in an UMP protostellar
disk with $[\mathrm{Fe/H}] \approx -4.1$ around the $M_{\ast} \approx
0.76~M_{\odot}$ star 2MASS J18082002--5104378 A can be used with Equations
(\ref{scaling1}) and (\ref{scaling2}) plus the data from \citet{tan14}
in Table~\ref{tbl-6} to explore the fragmentation mass scale and migration
times in primordial composition protostellar disks around Pop III stars.

According to \citet{tan14}, a protostellar disk around a $M_{\ast} =
1.0~M_{\odot}$ UMP star at $r = 10^{0.50}$ au is marginally stable
(i.e., $Q \approx 1$).  At $r \gtrsim 10^{1.0}$ au, that same disk
is gravitationally unstable (i.e., $Q < 1$).   We found that for a
$M_{\ast} = 10~M_{\odot}$ Pop III star, Equation (\ref{scaling1})
implies $M_{\mathrm{frag}} \approx (0.25, 0.88)~M_{\odot}$ at $r =
(10^{0.5},10^{1.0})$ au.  Let $t_{\mathrm{mig},1}$ be defined as the
migration time of a fragment in an UMP protostellar disk with the
same mass as 2MASS J18082002--5104378 B.  For that fragment at $r =
(10^{0.5},10^{1.0})$ au, Equation (\ref{scaling2}) indicates a migration
time $t_{\mathrm{mig},2} \approx (1.7, 1.4)~t_{\mathrm{mig},1}$.  For a
$M_{\ast} = 100~M_{\odot}$ Pop III star, the expected fragment mass is
$M_{\mathrm{frag}} \approx (1.0, 0.25)~M_{\odot}$ and $t_{\mathrm{mig},2}
\approx (22, 3.5)~t_{\mathrm{mig},1}$ at $r = (10^{0.5},10^{1})$ au.
In words, scaling from the 2MASS J18082002--5104378 system yields
fragments in the Pop III disk with $M_{\mathrm{frag}} \lesssim
1.0~M_{\odot}$ and $t_{\mathrm{mig},2} \gtrsim t_{\mathrm{mig},1}$.

We note that \citet{tan14} were forced to assume a \citet{sha73} $\alpha$
parameter to account for the efficiency of angular momentum transport
in their disk model.  In a disk with $Q \sim 1$, the angular momentum
transport parameterized by $\alpha$ is dominated by torques due to
spiral arms arising because the disk is near gravitational instability.
In this context, the contribution from these spiral arms parameterized by
$\alpha_{\mathrm{GI,max}}$ dominates the total angular momentum transport
compared to other sources like the magnetorotational instability.
While primordial composition disks in the fiducial model of \citet{tan14}
with $\alpha_{\mathrm{GI,max}} = 1$ will have $Q \approx 1.3$ at $a
= (10^{0.5},10^{1.0})$ au and therefore be formally stable against
fragmentation, they admit that lower values of $\alpha_{\mathrm{GI,max}}
= 0.07$ are equally plausible.  A smaller $\alpha_{\mathrm{GI,max}}$
would make a disk less able to transport angular momentum and more likely
to fragment.  Indeed, when $Q \sim 1$ Equation (15) of \citet{tan14}
shows that $Q$ is linearly proportional to $\alpha_{\mathrm{GI,max}}$.
Since the \citet{tan14} UMP and primordial composition disks have $Q
\approx 1.3$ when $\alpha_{\mathrm{GI,max}} = 1$, even a slightly smaller
$\alpha_{\mathrm{GI,max}} \approx 0.8$ would cause $Q$ to dip below 1
and therefore cause disk fragmentation.  For these reasons, we assert
that it is reasonable to consider the possibility that both primordial
composition and UMP disks will fragment at $a = (10^{0.5},10^{1.0})$ au.

\citet{han15} discovered an even lower mass companion in the EMP
single-lined spectroscopic binary system HE 1523--0901.  That system has
Keplerian orbital parameters $P = 303.05 \pm 0.25$ day, $K = 0.350 \pm
0.003$ km s$^{-1}$, and $e = 0.163 \pm 0.010$.  We calculated the mass of
the primary in the system---HE 1523--0901 A---as described in Section 3.2.
We found that it has $M_{\ast} = 0.83 \pm 0.01~M_{\odot}$.  The system's
Keplerian parameters and primary mass imply that the secondary in the
system---HE 1523--0901 B---has $M_{2} = 0.011_{-0.001}^{+0.003}~M_{\odot}
= 11_{-1}^{+4}~M_{\mathrm{Jup}}$.

At the edge of a protostellar disk where it is most unstable and
fragmentation is most likely, the properties of HE 1523--0901 B imply
fragmentation and fragment survival in both primordial composition disks
around the $M_{\ast} = 10~M_{\odot}$ and $M_{\ast} = 100~M_{\odot}$
Pop III stars considered above.  Let $t_{\mathrm{mig},1}$ be defined
as the migration time of a fragment in an UMP protostellar disk
with the same mass as HE 1523--0901 B starting from the edge of
that disk at $r = 10^{1.5}$ au.  For the edge of the disk around
the $M_{\ast} = 10~M_{\odot}$ star at $r = 10^{2.1}$ au, we found
$M_{\mathrm{frag}} \approx 0.18~M_{\odot}$ and $t_{\mathrm{mig},2}
\approx 12~t_{\mathrm{mig},1}$.  For the edge of the disk around
the $M_{\ast} = 100~M_{\odot}$ star at $r = 10^{3.2}$ au, we found
$M_{\mathrm{frag}} \approx 1.6~M_{\odot}$ and $t_{\mathrm{mig},2} \approx
52~t_{\mathrm{mig},1}$.  Once again, scaling from the HE 1523--0901
system yields fragments in the Pop III disk with $M_{\mathrm{frag}}
\sim 1.0~M_{\odot}$ and $t_{\mathrm{mig},2} \gtrsim t_{\mathrm{mig},1}$.
These calculations collectively support the idea that solar-mass fragments
can form in primordial composition disks around Pop III stars in the
mass range $10~M_{\odot} \lesssim M_{\ast} \lesssim 100~M_{\odot}$ and
subsequently survive disk migration without merging with the primary
forming at the center of the disk.

It is important to note several caveats to our inferences based on the
existence of 2MASS J18082002--5104378 B and HE 1523--0901 B.  If the 2MASS
J18082002--5104378 and HE 1523--0901 systems formed via fragmentation
at the filament or molecular core scales \citep[e.g.,][]{tur09,chi16},
then the scaling relation argument presented above would not apply.
Our study relies on the fidelity of the \citet{tan14} disk models, so
any unaccounted-for physics in those models will affect our analysis.
For example, it is possible that the transport of angular momentum in
UMP and primordial composition disks differs in a way that was not
accounted for by \citet{tan14}.  If that is so, then our scaling
relation analysis will be affected.  While our scaling relation
analysis does not rely on the absolute value of the fragment mass or
migration time predicted by Equations (\ref{mfrag}) and (\ref{tmig}), the
straightforward use of Equation (\ref{mfrag}) indicates a fragmentation
mass scale $M_{\mathrm{frag}} \sim 10^{-3}-10^{-4}~M_{\odot}$.  This is
significantly below the mass of 2MASS J18082002--5104378 B, and the
implication is that subsequent accretion was also important for its
formation.  Similarly, a simple application of Equation (\ref{tmig})
suggests a migration time that is an order of magnitude longer than
that observed in the detailed hydrodynamic models of \citet{hir17}.
Nevertheless, we argue that the advantage of a scaling analysis is that
it bypasses these apparent normalization issues.

\section{Discussion}

2MASS J18082002--5104378 B is significantly lower in mass than all other
known UMP stars.  Because most metal-poor star surveys are effectively
magnitude limited, most known metal-poor star are giants, subgiants,
or main sequence turnoff stars.  Few are low-mass main sequence
stars.  Of all stars with $[\mathrm{M/H}] \leq -2.5$ in the JINAbase
\citep{abo17} and the Stellar Abundances for Galactic Archaeology
\citep[SAGA;][]{sud08,sud11,yam13} databases, only 5/981 and 13/1314
have $\log{g} \geq 4.0$ and $T_{\mathrm{eff}} \leq 5000$ K.  The MIST
isochrones indicate that at its mass, 2MASS J18082002--5104378 B has
$T_{\mathrm{eff}} \approx 4200$ K, $\log{g} \approx 5.2$, bolometric
luminosity $L \approx 6.7 \times 10^{-3}~L_{\odot}$, and $V$-band
absolute magnitude $M_{V} \approx 10.8$.  At the distance of the 2MASS
J18082002--5104378 system, it likely has an unreddened $V \approx 19.7$.
Indeed, if 2MASS J18082002--5104378 B were a field star its ordinary
proper motion and faint apparent magnitude would never attract attention.

\begin{figure*}
\plottwo{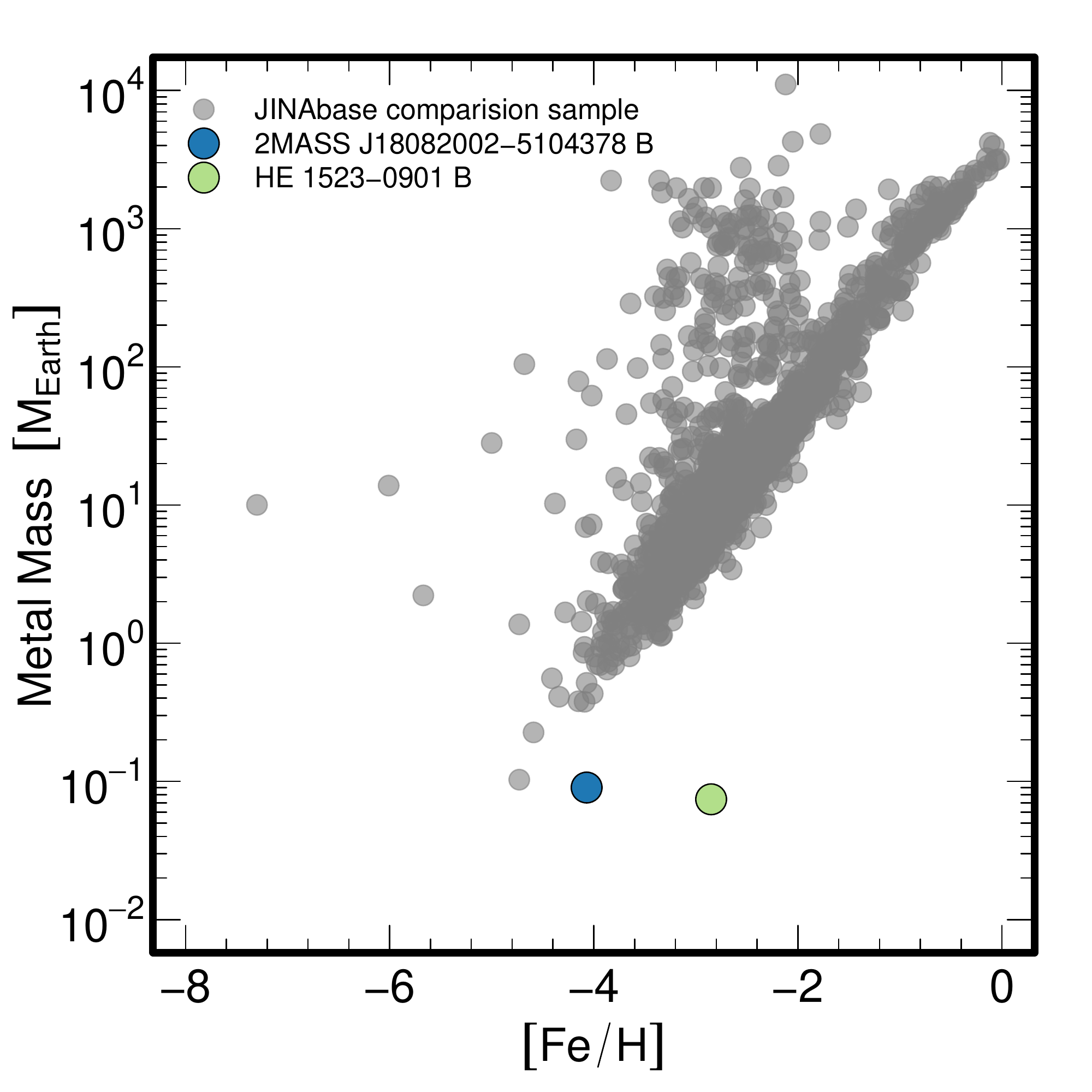}{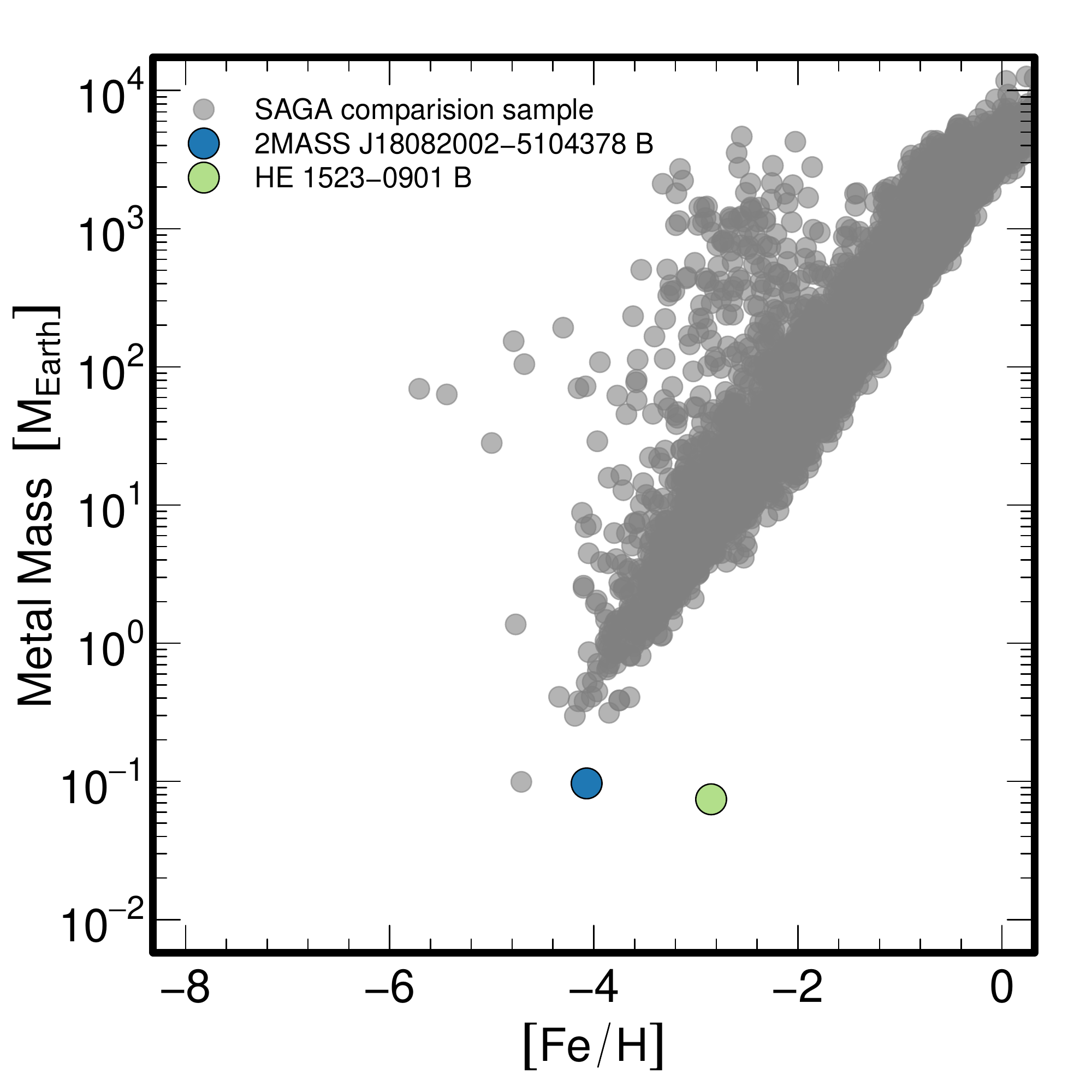}
\caption{Total metal mass as a function of iron metallicity for
known metal-poor stars.  Left: comparison with the JINAbase database.
Right: comparison with the SAGA database.  2MASS J18082002--5104378
B has a total metallicity $[\mathrm{M/H}] \approx -3.84$ and a
mass $M_{2} = 0.14_{-0.01}^{+0.06}~M_{\odot}$.  Because of its low
metallicity and mass, 2MASS J18082002--5104378 B has a total metal
mass of only $0.090~M_{\oplus} \approx 0.84~M_{\mathrm{Mars}}$,
making it the most metal-poor star known.  HE 1523--0901 B has a
total metallicity $[\mathrm{M/H}] \approx -2.81$ and a mass $M_{2} =
0.011_{-0.001}^{+0.003}~M_{\odot} = 11_{-1}^{+4}~M_{\mathrm{Jup}}$.
Because of its low metallicity and mass, HE 1523--0901 B has a total
metal mass of only $0.07~M_{\oplus} \approx 0.7~M_{\mathrm{Mars}}$.
The uncertainties on the total metal masses of 2MASS J18082002--5104378
B and HE 1523--0901 B are smaller than the plotted points.\label{fig05}}
\end{figure*}

Measured by its total mass in heavy elements, 2MASS J18082002--5104378
B is the most metal-poor star ever discovered.  We calculated the total
heavy element masses of 2MASS J18082002--5104378 B, HE 1523--0901 B,
and every star in the JINAbase and SAGA databases.  We focused on
the observable elements with logarithmic abundances $\epsilon > 6$
according to \citet{asp09}: carbon, nitrogen, oxygen, sodium, magnesium,
aluminum, silicon, sulfur, calcium, iron, and nickel.  Excluding the
unobservable noble gases neon and argon, together these 11 elements
comprise more than 99\% of the metal mass of a solar-composition star.
We assumed the standard atomic weights for each element and that each
star in the JINAbase and SAGA samples has $M_{\ast} = 0.8~M_{\odot}$.
We excluded upper limits and ignored missing elements when we summed the
total metal mass of a star.  We found that 2MASS J18082002--5104378 B has
total metal mass of $0.090~M_{\oplus} \approx 0.84~M_{\mathrm{Mars}}$.
It has fewer grams of heavy elements than the next closest star SDSS
J102915.14+172927.9 \citep{caf11,caf12}, which has a total metal mass
of $0.10~M_{\oplus} \approx 1.1~M_{\mathrm{Mars}}$.  The brown dwarf
HE 1523--0901 B is even more extreme, as it has a total metal mass
of only $0.07~M_{\oplus} \approx 0.7~M_{\mathrm{Mars}}$.  We plot in
Figure~\ref{fig05} total metal mass as a function of [Fe/H] for 2MASS
J18082002--5104378 B, HE 1523--0901 B, and every star in the JINAbase
and SAGA databases.

The 2MASS J18082002--5104378 system is the most metal-poor star system
on a thin disk orbit yet found.  Its thin disk orbit is unusual for a
metal-poor star, as only one out of 101 stars from \citet{bee17} with
$\mathrm{[Fe/H]} \lesssim -2.5$ have similar Galactic orbital parameters.
One problem with this simple analysis is that most searches for metal-poor
stars avoid the plane of the Galaxy, so observational selection effects
may have made it very difficult to discover similarly metal-poor stars
with thin disk orbits in the past.

Given its thin disk orbit, the $13.535 \pm 0.002$ Gyr age of the 2MASS
J18082002--5104378 system provides a lower limit on the age of the
thin disk.  Similarly old but not quite as metal-poor stars have also
been seen on thin disk orbits \citep[e.g.,][]{cas11,ben14}.  This is
somewhat older than the 8--10 Gyr age of the thin disk suggested by
classical studies of field stars \citep[][]{edv93,liu00,san03}, the
white dwarf luminosity function \citep[e.g.,][]{osw96,leg98,kno99,kil17},
and the ages of the oldest disk open clusters Berkeley 17 and NGC 6791
\citep[e.g.,][]{kru06,bro12}.

While 2MASS J18082002--5104378 A is a subgiant and therefore likely to
yield a reasonable isochrone-derived mass and age, our random mass and
age uncertainties are probably too small.  Both are almost certainly
affected by significant systematic uncertainties.  For example,
the Dartmouth isochrone grid we fit to our observational data has
not accounted for the possible non-solar carbon abundance of the
2MASS J18082002--5104378 system or the possibility that important
stellar model parameters like mixing length may vary with composition
\citep[e.g.,][]{bon12,met14,tay17,joy18,via18}.  Nevertheless, the low
metallicity and ancient age of 2MASS J18082002--5104378 does suggest
that the thin disk may be older than usually assumed.

Our conclusions about the likely formation of low-mass stars via
disk fragmentation in primordial composition protostellar disks
around Pop III stars support the fidelity of disk fragmentation
events seen in recent numerical simulations of Pop III star formation
\citep[e.g.,][]{cla08,cla11a,cla11b,sta10,sta12,sta16,gre11,gre12,dop13,sta13,sta14,hir17,ria18}.
At the resolutions necessary to observe fragmentation, these simulations
can only be run for order 10 dynamical times, far too short a time to
follow any possible migration of a newly formed fragment due to disk
torques and evaluate its survival.  Our scaling relation analysis of
the 2MASS J18082002--5104378 and HE 1523--0901 systems implied that
the migration times of fragments in Pop III disks with appropriately
scaled masses should be longer than the disk migration times of 2MASS
J18082002--5104378 B and HE 1523--0901 B.  Therefore, since they both
survived, fragments in primordial composition disks with migration
times even longer could survive as well.  We therefore argue that disk
fragmentation is likely to form low-mass Pop III stars.  We further
suggest that at least some of those fragments are likely to survive as
low-mass stars instead of merging with the primary forming at the center
of their stellar system.  These surviving low-mass Pop III stars would
have main sequence lifetimes long enough that they could persist in our
Galaxy to the present day.

\section{Conclusion}

We report the discovery of a low-mass secondary star in the 13.5 Gyr old,
$[\mathrm{Fe/H}] \approx -4.1$ single-lined spectroscopic binary system
2MASS J18082002--5104378.  The secondary star 2MASS J18082002--5104378
B has a mass $M_{2} = 0.14_{-0.01}^{+0.06}~M_{\odot}$, very close to the
hydrogen-burning limit for its composition.  It is the lowest-mass ultra
metal-poor star currently known.  Because of its low mass and metallicity,
2MASS J18082002--5104378 B has fewer grams of heavy elements than any
other star currently known.  Despite its age and metallicity, the 2MASS
J18082002--5104378 system is on a thin disk like orbit.  Indeed, it is the
most metal-poor star system yet found to be kinematically associated with
the thin disk.  In concert with theoretical models of protostellar disks
around both ultra metal-poor and primordial composition protostars, the
observed properties of 2MASS J18082002--5104378 B and the $[\mathrm{Fe/H}]
\sim -3$ brown dwarf HE 1523--0901 B support the theoretically proposed
idea that low-mass Pop III stars form via disk fragmentation.  While
low-mass Pop III star formation via disk fragmentation has been seen in
numerical simulations, it is impossible to run those simulations long to
enough to verify that fragments survive disk migration.  Our discovery
reveals for the first time that fragments do survive the era of disk
migration.  This inference is independent of poorly modeled supernova
yields or the possible atmospheric contamination of Pop III stars, two
issues that have hindered past observational efforts to validate the
existence of low-mass Pop III stars.  Collectively, these results imply
that low-mass Pop III stars formed via disk fragmentation can exist in
our Galaxy.

\acknowledgments
We thank David Nataf, Brian O'Shea, and Jason Tumlinson for helpful
comments.  We are grateful to the anonymous referee for providing
an extraordinarily quick and insightful referee report with several
suggestions that improved the paper.  A.R.C. is supported through an
Australian Research Council Discovery Project under grant DP160100637.
This paper includes data gathered with the 6.5 m Magellan Telescopes
located at Las Campanas Observatory, Chile.  Based on data obtained from
the ESO Science Archive Facility under request number 270734.  Based on
observations collected at the European Organisation for Astronomical
Research in the Southern Hemisphere under ESO programme 293.D-5036(A).
Based on observations obtained under programs GS-2016A-Q-76,
GS-2016B-Q-80, and GS-2017A-Q-66 at the Gemini Observatory, which is
operated by the Association of Universities for Research in Astronomy,
Inc., under a cooperative agreement with the NSF on behalf of the
Gemini partnership: the National Science Foundation (United States),
the National Research Council (Canada), CONICYT (Chile), Ministerio
de Ciencia, Tecnolog\'{i}a e Innovaci\'{o}n Productiva (Argentina),
and Minist\'{e}rio da Ci\^{e}ncia, Tecnologia e Inova\c{c}\~{a}o
(Brazil).  This work has made use of data from the European Space Agency
(ESA) mission \textit{Gaia} (\url{https://www.cosmos.esa.int/gaia}),
processed by the \textit{Gaia} Data Processing and Analysis Consortium
(DPAC, \url{https://www.cosmos.esa.int/web/gaia/dpac/consortium}).
Funding for the DPAC has been provided by national institutions,
in particular the institutions participating in the \textit{Gaia}
Multilateral Agreement.  The national facility capability for SkyMapper
has been funded through ARC LIEF grant LE130100104 from the Australian
Research Council, awarded to the University of Sydney, the Australian
National University, Swinburne University of Technology, the University
of Queensland, the University of Western Australia, the University of
Melbourne, Curtin University of Technology, Monash University and the
Australian Astronomical Observatory.  SkyMapper is owned and operated
by The Australian National University's Research School of Astronomy
and Astrophysics.  The survey data were processed and provided by
the SkyMapper Team at ANU.  The SkyMapper node of the All-Sky Virtual
Observatory (ASVO) is hosted at the National Computational Infrastructure
(NCI).  Development and support the SkyMapper node of the ASVO has been
funded in part by Astronomy Australia Limited (AAL) and the Australian
Government through the Commonwealth's Education Investment Fund (EIF)
and National Collaborative Research Infrastructure Strategy (NCRIS),
particularly the National eResearch Collaboration Tools and Resources
(NeCTAR) and the Australian National Data Service Projects (ANDS).
This publication makes use of data products from the Two Micron All
Sky Survey, which is a joint project of the University of Massachusetts
and the Infrared Processing and Analysis Center/California Institute of
Technology, funded by the National Aeronautics and Space Administration
and the National Science Foundation.  This publication makes use of data
products from the \textit{Wide-field Infrared Survey Explorer}, which
is a joint project of the University of California, Los Angeles, and the
Jet Propulsion Laboratory/California Institute of Technology, funded by
the National Aeronautics and Space Administration.  This research was
made possible through the use of the AAVSO Photometric All-Sky Survey
(APASS), funded by the Robert Martin Ayers Sciences Fund.  This research
has made use of NASA's Astrophysics Data System Bibliographic Services.
This research made use of Astropy, a community-developed core Python
package for Astronomy \citep{ast13,ast18}.  This research has made use of
the SIMBAD database, operated at CDS, Strasbourg, France \citep{wen00}.
This research has made use of the VizieR catalogue access tool, CDS,
Strasbourg, France.  The original description of the VizieR service was
published in A\&AS 143, 23 \citep{och00}.

\vspace{5mm}
\facilities{Magellan:Clay (MIKE echelle spectrograph),
            VLT:Kueyen (UVES echelle spectrograph),
            Gemini:South (GMOS-S imaging spectrograph).}

\software{\texttt{astropy} \citep{ast13,ast18},
          \texttt{CarPy} \citep{kel00,kel03},
          \texttt{ExoFit} \citep{bal09},
          \texttt{galpy} \citep{bov15},
          \texttt{isochrones} \citep{mor15},
          \texttt{numpy} \citep{oli06},
          \texttt{MultiNest} \citep{fer08,fer09,fer13},
          \texttt{pandas} \citep{mck10},
          \texttt{R} \citep{r18},
          \texttt{scipy} \citep{joh01}
          }



\begin{thebibliography}{}
\bibitem[Abel et al.(2000)]{abe00} Abel, T., Bryan, G.~L., \&
Norman, M.~L.\ 2000, \apj, 540, 39
\bibitem[Abel et al.(2002)]{abe02} Abel, T., Bryan, G.~L., \&
Norman, M.~L.\ 2002, Science, 295, 93
\bibitem[Abohalima \& Frebel(2017)]{abo17} Abohalima, A., \& Frebel, A.\
2017, arXiv:1711.04410
\bibitem[Alpher et al.(1948)]{alp48} Alpher, R.~A., Bethe, H., \& Gamow, G.\
1948, Physical Review, 73, 803
\bibitem[Aoki et al.(2014)]{aok14} Aoki, W., Tominaga, N., Beers, T.~C.,
Honda, S., \& Lee, Y.~S.\ 2014, Science, 345, 912
\bibitem[Arenou et al.(2018)]{are18} Arenou, F., Luri, X., Babusiaux, C.,
et al.\ 2018, \aap, 616, A17 
\bibitem[Asplund et al.(2009)]{asp09} Asplund, M., Grevesse, N.,
Sauval, A.~J., \& Scott, P.\ 2009, \araa, 47, 481
\bibitem[Astropy Collaboration et al.(2018)]{ast18} Astropy Collaboration,
Price-Whelan, A.~M., Sip{\H o}cz, B.~M., et al.\ 2018, \aj, 156, 123   
\bibitem[Astropy Collaboration et al.(2013)]{ast13} Astropy Collaboration,
Robitaille, T.~P., Tollerud, E.~J., et al.\ 2013, \aap, 558, A33
\bibitem[Balan \& Lahav(2009)]{bal09} Balan, S.~T., \& Lahav, O.\
2009, \mnras, 394, 1936
\bibitem[Baruteau et al.(2011)]{bar11} Baruteau, C., Meru, F., \&
Paardekooper, S.-J.\ 2011, \mnras, 416, 1971
\bibitem[Beers \& Christlieb(2005)]{bee05} Beers, T.~C., \& Christlieb, N.\
2005, \araa, 43, 531
\bibitem[Beers et al.(2017)]{bee17} Beers, T.~C., Placco, V.~M.,
Carollo, D., et al.\ 2017, \apj, 835, 81
\bibitem[Bensby et al.(2014)]{ben14} Bensby, T., Feltzing, S., \&
Oey, M.~S.\ 2014, \aap, 562, A71
\bibitem[Bernstein et al.(2003)]{ber03} Bernstein, R., Shectman, S.~A.,
Gunnels, S.~M., Mochnacki, S., \& Athey, A.~E.\ 2003, \procspie, 4841, 1694
\bibitem[Bland-Hawthorn \& Gerhard(2016)]{bla16} Bland-Hawthorn, J., \&
Gerhard, O.\ 2016, \araa, 54, 529
\bibitem[Bonaca et al.(2012)]{bon12} Bonaca, A., Tanner, J.~D., Basu, S.,
et al.\ 2012, \apjl, 755, L12
\bibitem[Bovy(2015)]{bov15} Bovy, J.\ 2015, \apjs, 216, 29
\bibitem[Brogaard et al.(2012)]{bro12} Brogaard, K., VandenBerg, D.~A.,
Bruntt, H., et al.\ 2012, \aap, 543, A106
\bibitem[Bromm(2013)]{bro13} Bromm, V.\ 2013, Reports on Progress in Physics,
76, 112901
\bibitem[Bromm et al.(1999)]{bro99} Bromm, V., Coppi, P.~S., \&
Larson, R.~B.\ 1999, \apjl, 527, L5
\bibitem[Bromm et al.(2002)]{bro02} Bromm, V., Coppi, P.~S., \&
Larson, R.~B.\ 2002, \apj, 564, 23
\bibitem[Caffau et al.(2011)]{caf11} Caffau, E., Bonifacio, P.,
Fran{\c c}ois, P., et al.\ 2011, \nat, 477, 67
\bibitem[Caffau et al.(2012)]{caf12} Caffau, E., Bonifacio, P.,
Fran{\c c}ois, P., et al.\ 2012, \aap, 542, A51
\bibitem[Casagrande et al.(2011)]{cas11} Casagrande, L., Sch{\"o}nrich, R.,
Asplund, M., et al.\ 2011, \aap, 530, A138
\bibitem[Chiaki et al.(2016)]{chi16} Chiaki, G., Yoshida, N., \& Hirano, S.\
2016, \mnras, 463, 2781
\bibitem[Choi et al.(2016)]{cho16} Choi, J., Dotter, A., Conroy, C.,
et al.\ 2016, \apj, 823, 102
\bibitem[Clark et al.(2008)]{cla08} Clark, P.~C., Glover, S.~C.~O., \&
Klessen, R.~S.\ 2008, \apj, 672, 757
\bibitem[Clark et al.(2011a)]{cla11a} Clark, P.~C., Glover, S.~C.~O.,
Klessen, R.~S., \& Bromm, V.\ 2011a, \apj, 727, 110
\bibitem[Clark et al.(2011b)]{cla11b} Clark, P.~C., Glover, S.~C.~O.,
Smith, R.~J., et al.\ 2011b, Science, 331, 1040
\bibitem[Coelho et al.(2005)]{coe05} Coelho, P., Barbuy, B., Mel{\'e}ndez, J.,
Schiavon, R.~P., \& Castilho, B.~V.\ 2005, \aap, 443, 735
\bibitem[Cyburt et al.(2016)]{cyb16} Cyburt, R.~H., Fields, B.~D.,
Olive, K.~A., \& Yeh, T.-H.\ 2016, Reviews of Modern Physics, 88, 015004
\bibitem[da Silva et al.(2006)]{das06} da Silva, L., Girardi, L.,
Pasquini, L., et al.\ 2006, \aap, 458, 609
\bibitem[de Bennassuti et al.(2017)]{deb17} de Bennassuti, M., Salvadori, S.,
Schneider, R., Valiante, R., \& Omukai, K.\ 2017, \mnras, 465, 926
\bibitem[Dekker et al.(2000)]{dek00} Dekker, H., D'Odorico, S., Kaufer, A.,
Delabre, B., \& Kotzlowski, H.\ 2000, \procspie, 4008, 534
\bibitem[Dopcke et al.(2013)]{dop13} Dopcke, G., Glover, S.~C.~O.,
Clark, P.~C., \& Klessen, R.~S.\ 2013, \apj, 766, 103
\bibitem[Dotter(2016)]{dot16} Dotter, A.\ 2016, \apjs, 222, 8
\bibitem[Dotter et al.(2007)]{dot07} Dotter, A., Chaboyer, B.,
Jevremovi{\'c}, D., et al.\ 2007, \aj, 134, 376
\bibitem[Dotter et al.(2008)]{dot08} Dotter, A., Chaboyer, B.,
Jevremovi{\'c}, D., et al.\ 2008, \apjs, 178, 89
\bibitem[Edvardsson et al.(1993)]{edv93} Edvardsson, B., Andersen, J.,
Gustafsson, B., et al.\ 1993, \aap, 275, 101
\bibitem[Evans et al.(2018)]{eva18} Evans, D.~W., Riello, M., De Angeli, F.,
et al.\ 2018, \aap, 616, A4
\bibitem[Feroz \& Hobson(2008)]{fer08} Feroz, F., \& Hobson, M.~P.\ 2008,
\mnras, 384, 449
\bibitem[Feroz et al.(2009)]{fer09} Feroz, F., Hobson, M.~P., \& Bridges, M.\
2009, \mnras, 398, 1601
\bibitem[Feroz et al.(2013)]{fer13} Feroz, F., Hobson, M.~P., Cameron, E., \&
Pettitt, A.~N.\ 2013, arXiv:1306.2144
\bibitem[Fraser et al.(2017)]{fra17} Fraser, M., Casey, A.~R., Gilmore, G.,
Heger, A., \& Chan, C.\ 2017, \mnras, 468, 418
\bibitem[Frebel et al.(2009)]{fre09} Frebel, A., Johnson, J.~L., \&
Bromm, V.\ 2009, \mnras, 392, L50
\bibitem[Gaia Collaboration et al.(2018)]{gai18} Gaia Collaboration,
Brown, A.~G.~A., Vallenari, A., et al.\ 2018, \aap, 616, A1
\bibitem[Gaia Collaboration et al.(2016)]{gai16} Gaia Collaboration,
Prusti, T., de Bruijne, J.~H.~J., et al.\ 2016, \aap, 595, A1
\bibitem[Gimeno et al.(2016)]{gim16} Gimeno, G., Roth, K.,
Chiboucas, K., et al.\ 2016, \procspie, 9908, 99082S
\bibitem[Glover(2013)]{glo13} Glover, S.\ 2013, in The First Galaxies,
Astrophysics and Space Science Library, Vol. 396, ed. T. Wiklind,
B. Mobasher, \& V. Bromm (Berlin: Springer), 103
\bibitem[Gravity Collaboration et al.(2018)]{gra18} Gravity Collaboration,
Abuter, R., Amorim, A., et al.\ 2018, \aap, 615, L15
\bibitem[Greif(2015)]{gre15} Greif, T.~H.\ 2015, Computational Astrophysics
and Cosmology, 2, 3
\bibitem[Greif et al.(2012)]{gre12} Greif, T.~H., Bromm, V.,
Clark, P.~C., et al.\ 2012, \mnras, 424, 399
\bibitem[Greif et al.(2011)]{gre11} Greif, T.~H., Springel, V.,
White, S.~D.~M., et al.\ 2011, \apj, 737, 75
\bibitem[Halbwachs(1987)]{hal87} Halbwachs, J.~L.\ 1987, \aap, 183, 234
\bibitem[Hambly et al.(2018)]{ham18} Hambly, N.~C., Cropper, M.,
Boudreault, S., et al.\ 2018, \aap, 616, A15
\bibitem[Hansen et al.(2015)]{han15} Hansen, T.~T., Andersen, J.,
Nordstr{\"o}m, B., et al.\ 2015, \aap, 583, A49
\bibitem[Hartwig et al.(2015)]{har15} Hartwig, T., Bromm, V.,
Klessen, R.~S., \& Glover, S.~C.~O.\ 2015, \mnras, 447, 3892
\bibitem[Henden et al.(2016)]{hen16} Henden, A.~A., Templeton, M.,
Terrell, D., et al.\ 2016, yCat, 2336, 0
\bibitem[Hirano \& Bromm(2017)]{hir17} Hirano, S., \& Bromm, V.\ 2017, \mnras,
470, 898
\bibitem[Hirano et al.(2015)]{hir15} Hirano, S., Hosokawa, T., Yoshida, N.,
Omukai, K., \& Yorke, H.~W.\ 2015, \mnras, 448, 568 
\bibitem[Hirano et al.(2014)]{hir14} Hirano, S., Hosokawa, T., Yoshida, N.,
et al.\ 2014, \apj, 781, 60
\bibitem[Hogeveen(1992)]{hog92} Hogeveen, S.~J.\ 1992, \apss, 194, 143
\bibitem[Hook et al.(2004)]{hoo04} Hook, I.~M., J{\o}rgensen, I.,
Allington-Smith, J.~R., et al.\ 2004, \pasp, 116, 425
\bibitem[Hosokawa et al.(2016)]{hos16} Hosokawa, T., Hirano, S.,
Kuiper, R., et al.\ 2016, \apj, 824, 119
\bibitem[Hosokawa et al.(2011)]{hos11} Hosokawa, T., Omukai, K.,
Yoshida, N., \& Yorke, H.~W.\ 2011, Science, 334, 1250
\bibitem[Ishigaki et al.(2018)]{ish18} Ishigaki, M.~N., Tominaga, N.,
Kobayashi, C., \& Nomoto, K.\ 2018, \apj, 857, 46
\bibitem[Johnson(2015)]{joh15} Johnson, J.~L.\ 2015, \mnras, 453, 2771
\bibitem[Jones et al.(2001)]{joh01} Jones, E., Oliphant, E., \& Peterson, P.,
SciPy: Open Source Scientific Tools for Python, 2001,
\url{http://www.scipy.org/}
\bibitem[Joyce \& Chaboyer(2018)]{joy18} Joyce, M., \& Chaboyer, B.\ 2018,
\apj, 856, 10
\bibitem[Juri{\'c} et al.(2008)]{jur08} Juri{\'c}, M., Ivezi{\'c}, {\v Z}.,
Brooks, A., et al.\ 2008, \apj, 673, 864
\bibitem[Kelson(2003)]{kel03} Kelson, D.~D.\ 2003, \pasp, 115, 688
\bibitem[Kelson et al.(2000)]{kel00} Kelson, D.~D., Illingworth, G.~D.,
van Dokkum, P.~G., \& Franx, M.\ 2000, \apj, 531, 137
\bibitem[Kilic et al.(2017)]{kil17} Kilic, M., Munn, J.~A.,
Harris, H.~C., et al.\ 2017, \apj, 837, 162
\bibitem[Komiya et al.(2015)]{kom15} Komiya, Y., Suda, T., \&
Fujimoto, M.~Y.\ 2015, \apjl, 808, L47
\bibitem[Komiya et al.(2016)]{kom16} Komiya, Y., Suda, T., \&
Fujimoto, M.~Y.\ 2016, \apj, 820, 59
\bibitem[Knox et al.(1999)]{kno99} Knox, R.~A., Hawkins, M.~R.~S., \&
Hambly, N.~C.\ 1999, \mnras, 306, 736
\bibitem[Kratter \& Lodato(2016)]{kra16} Kratter, K., \& Lodato, G.\ 2016,
\araa, 54, 271
\bibitem[Krusberg \& Chaboyer(2006)]{kru06} Krusberg, Z.~A.~C., \&
Chaboyer, B.\ 2006, \aj, 131, 1565
\bibitem[Leggett et al.(1998)]{leg98} Leggett, S.~K., Ruiz, M.~T., \&
Bergeron, P.\ 1998, \apj, 497, 294
\bibitem[Lindegren et al.(2018)]{lin18} Lindegren, L., Hern{\'a}ndez, J.,
Bombrun, A., et al.\ 2018, \aap, 616, A2
\bibitem[Liu \& Chaboyer(2000)]{liu00} Liu, W.~M., \& Chaboyer, B.\ 2000, \apj
544, 818
\bibitem[Luri et al.(2018)]{lur18} Luri, X., Brown, A.~G.~A., Sarro, L.~M.,
et al.\ 2018, \aap, 616, A9
\bibitem[Mainzer et al.(2011)]{mai11} Mainzer, A., Bauer, J., Grav, T.,
et al.\ 2011, \apj, 731, 53
\bibitem[McKee \& Tan(2008)]{mck08} McKee, C.~F., \& Tan, J.~C.\ 2008, \apj,
681, 771
\bibitem[McKinney(2010)]{mck10} McKinney, W.\ 2010, Proc. IX Python in
Science Conf., ed. S. van der Walt \& J. Millman (Austin, TX: SciPy),
51, \url{http://conference.scipy.org/proceedings/scipy2010/}
\bibitem[Mel{\'e}ndez et al.(2016)]{mel16} Mel{\'e}ndez, J., Placco, V.~M.,
Tucci-Maia, M., et al.\ 2016, \aap, 585, L5
\bibitem[Metcalfe et al.(2014)]{met14} Metcalfe, T.~S., Creevey, O.~L.,
Do{\u g}an, G., et al.\ 2014, \apjs, 214, 27
\bibitem[Miyamoto \& Nagai(1975)]{miy75} Miyamoto, M., \& Nagai, R.\
1975, \pasj, 27, 533
\bibitem[Morton(2015)]{mor15} Morton, T.~D.\ 2015, Isochrones: Stellar
Model Grid Package, Astrophysics Source Code Library, ascl:1503.010
\bibitem[Navarro et al.(1996)]{nav96} Navarro, J.~F., Frenk, C.~S., \&
White, S.~D.~M.\ 1996, \apj, 462, 563
\bibitem[O'Shea \& Norman(2007)]{osh07} O'Shea, B.~W., \& Norman, M.~L.\
2007, \apj, 654, 66
\bibitem[Ochsenbein et al.(2000)]{och00} Ochsenbein, F., Bauer, P., \&
Marcout, J.\ 2000, \aaps, 143, 23
\bibitem[Oliphant(2006)]{oli06} Oliphant, T.~E.\ 2006, A Guide to NumPy
(Spanish Fork, UT: Trelgol Publishing)
\bibitem[Oswalt et al.(1996)]{osw96} Oswalt, T.~D., Smith, J.~A.,
Wood, M.~A., \& Hintzen, P.\ 1996, \nat, 382, 692
\bibitem[Paardekooper \& Johansen(2018)]{paa18} Paardekooper, S.-J., \&
Johansen, A.\ 2018, \ssr, 214, 38
\bibitem[Paxton et al.(2011)]{pax11} Paxton, B., Bildsten, L., Dotter, A.,
et al.\ 2011, \apjs, 192, 3
\bibitem[Paxton et al.(2013)]{pax13} Paxton, B., Cantiello, M., Arras, P.,
et al.\ 2013, \apjs, 208, 4
\bibitem[Paxton et al.(2015)]{pax15} Paxton, B., Marchant, P., Schwab, J.,
et al.\ 2015, \apjs, 220, 15
\bibitem[Placco et al.(2015)]{pla15} Placco, V.~M., Frebel, A., Lee, Y.~S.,
et al.\ 2015, \apj, 809, 136
\bibitem[R Core Team(2018)]{r18} R Core Team 2018, R: A Language and
Environment for Statistical Computing (Vienna: R Foundation for
Statistical Computing)
\bibitem[Riaz et al.(2018)]{ria18} Riaz, R., Bovino, S., Vanaverbeke, S., \&
Schleicher, D.~R.~G.\ 2018, \mnras, 479, 667
\bibitem[Riello et al.(2018)]{rie18} Riello, M., De Angeli, F., Evans, D.~W.,
et al.\ 2018, \aap, 616, A3
\bibitem[Salgado et al.(2017)]{sal17} Salgado, J.,
Gonz{\'a}lez-N{\'u}{\~n}ez, J., Guti{\'e}rrez-S{\'a}nchez, R., et al.\ 2017,
Astronomy and Computing, 21, 22
\bibitem[Salvadori et al.(2007)]{sal07} Salvadori, S., Schneider, R., \&
Ferrara, A.\ 2007, \mnras, 381, 647
\bibitem[Sandage et al.(2003)]{san03} Sandage, A., Lubin, L.~M., \&
VandenBerg, D.~A.\ 2003, \pasp, 115, 1187
\bibitem[Saumon et al.(1994)]{sau94} Saumon, D., Bergeron, P., Lunine, J.~I.,
Hubbard, W.~B., \& Burrows, A.\ 1994, \apj, 424, 333
\bibitem[Schneider et al.(2012)]{sch12} Schneider, R., Omukai, K.,
Bianchi, S., \& Valiante, R.\ 2012, \mnras, 419, 1566
\bibitem[Shakura \& Sunyaev(1973)]{sha73} Shakura, N.~I., \& Sunyaev, R.~A.\
1973, \aap, 24, 337
\bibitem[Shectman \& Johns(2003)]{she03} Shectman, S.~A., \& Johns, M.\ 2003,
\procspie, 4837, 910
\bibitem[Shen et al.(2017)]{she17} Shen, S., Kulkarni, G., Madau, P., \&
Mayer, L.\ 2017, \mnras, 469, 4012
\bibitem[Silk(1983)]{sil83} Silk, J.\ 1983, \mnras, 205, 705
\bibitem[Skrutskie et al.(2006)]{skr06} Skrutskie, M.~F., Cutri, R.~M.,
Stiening, R., et al.\ 2006, \aj, 131, 1163
\bibitem[Soubiran et al.(2013)]{sou13} Soubiran, C., Jasniewicz, G.,
Chemin, L., et al.\ 2013, \aap, 552, A64
\bibitem[Stacy \& Bromm(2013)]{sta13} Stacy, A., \& Bromm, V.\ 2013, \mnras,
433, 1094
\bibitem[Stacy \& Bromm(2014)]{sta14} Stacy, A., \& Bromm, V.\ 2014, \apj, 785,
73
\bibitem[Stacy et al.(2016)]{sta16} Stacy, A., Bromm, V., \& Lee, A.~T.\ 2016,
\mnras, 462, 1307
\bibitem[Stacy et al.(2010)]{sta10} Stacy, A., Greif, T.~H., \&
Bromm, V.\ 2010, \mnras, 403, 45
\bibitem[Stacy et al.(2012)]{sta12} Stacy, A., Greif, T.~H., \&
Bromm, V.\ 2012, \mnras, 422, 290
\bibitem[Suda et al.(2008)]{sud08} Suda, T., Katsuta, Y.,
Yamada, S., et al.\ 2008, \pasj, 60, 1159
\bibitem[Suda et al.(2011)]{sud11} Suda, T., Yamada, S.,
Katsuta, Y., et al.\ 2011, \mnras, 412, 843
\bibitem[Susa(2013)]{sus13} Susa, H.\ 2013, \apj, 773, 185
\bibitem[Susa et al.(2014)]{sus14} Susa, H., Hasegawa, K., \&
Tominaga, N.\ 2014, \apj, 792, 32
\bibitem[Takahashi et al.(2014)]{tak14} Takahashi, K., Umeda, H., \&
Yoshida, T.\ 2014, \apj, 794, 40
\bibitem[Takahashi et al.(2018)]{tak18} Takahashi, K., Yoshida, T., \&
Umeda, H.\ 2018, \apj, 857, 111
\bibitem[Tanaka \& Omukai(2014)]{tan14} Tanaka, K.~E.~I., \& Omukai, K.\ 2014,
\mnras, 439, 1884
\bibitem[Tanaka et al.(2017)]{tan17} Tanaka, S.~J., Chiaki, G., Tominaga, N.,
\& Susa, H.\ 2017, \apj, 844, 137
\bibitem[Tanikawa et al.(2018)]{tan18} Tanikawa, A., Suzuki, T.~K., \&
Doi, Y.\ 2018, \pasj, 70, 80
\bibitem[Tayar et al.(2017)]{tay17} Tayar, J., Somers, G.,
Pinsonneault, M.~H., et al.\ 2017, \apj, 840, 17
\bibitem[Tegmark et al.(1997)]{teg97} Tegmark, M., Silk, J., Rees, M.~J.,
et al.\ 1997, \apj, 474, 1
\bibitem[Tody(1986)]{tod86} Tody, D.\ 1986, \procspie, 627, 733
\bibitem[Tody(1993)]{tod93} Tody, D.\ 1993, adass II, 52, 173
\bibitem[Tominaga et al.(2014)]{tom14} Tominaga, N., Iwamoto, N., \&
Nomoto, K.\ 2014, \apj, 785, 98
\bibitem[Tonry \& Davis(1979)]{ton79} Tonry, J., \& Davis, M.\ 1979, \aj, 84,
1511
\bibitem[Tumlinson(2006)]{tum06} Tumlinson, J.\ 2006, \apj, 641, 1
\bibitem[Turk et al.(2009)]{tur09} Turk, M.~J., Abel, T., \&
O'Shea, B.\ 2009, Science, 325, 601
\bibitem[Viani et al.(2018)]{via18} Viani, L.~S., Basu, S., Joel Ong J., M.,
Bonaca, A., \& Chaplin, W.~J.\ 2018, \apj, 858, 28
\bibitem[Vorobyov et al.(2013)]{vor13} Vorobyov, E.~I., DeSouza, A.~L., \&
Basu, S.\ 2013, \apj, 768, 131
\bibitem[Wenger et al.(2000)]{wen00} Wenger, M., Ochsenbein, F., Egret, D.,
et al.\ 2000, \aaps, 143, 9
\bibitem[Wolf et al.(2018)]{wol18} Wolf, C., Onken, C.~A., Luvaul, L.~C.,
et al.\ 2018, \pasa, 35, e010
\bibitem[Wright et al.(2010)]{wri10} Wright, E.~L., Eisenhardt, P.~R.~M.,
Mainzer, A.~K., et al.\ 2010, \aj, 140, 1868
\bibitem[Yamada et al.(2013)]{yam13} Yamada, S., Suda, T., Komiya, Y.,
Aoki, W., \& Fujimoto, M.~Y.\ 2013, \mnras, 436, 1362
\bibitem[Yoshida et al.(2006)]{yos06} Yoshida, N., Omukai, K.,
Hernquist, L., \& Abel, T.\ 2006, \apj, 652, 6
\end{thebibliography}
\end{document}